\documentclass{article}

\usepackage{arxiv}

\usepackage[utf8]{inputenc} 
\usepackage[T1]{fontenc}    
\usepackage{hyperref}       
\usepackage{url}            
\usepackage{booktabs}       
\usepackage{amsfonts}       
\usepackage{nicefrac}       
\usepackage{microtype}      
\usepackage{lipsum}
\usepackage{graphicx}
\graphicspath{ {./images/} }

\usepackage{amsfonts}   
\usepackage{amssymb}    
\usepackage{amsmath}
\usepackage[table]{xcolor}

\usepackage{cleveref}
\crefrangelabelformat{equation}{(#3#1#4--#5#2#6)}
\crefname{equation}{Eq.}{Eqs.}
\Crefname{equation}{Equation}{Equations}

\usepackage{amsthm}
\usepackage{mathtools}
\usepackage{amsfonts} 

\usepackage{arxiv}
\usepackage{algorithm, algorithmic}

\usepackage[nocomma]{optidef}

\usepackage{empheq}
\usepackage{graphicx}

\usepackage[utf8]{inputenc} 
\usepackage[T1]{fontenc}    
\usepackage{hyperref}       
\usepackage{url}            
\usepackage{amsfonts}       
\usepackage{booktabs}       
\usepackage{nicefrac}       
\usepackage{microtype}      
\usepackage{lipsum}
\usepackage{graphicx}
\graphicspath{ {./images/} }
\usepackage{float}
\usepackage{caption}
\usepackage[font=small,labelfont=bf, labelsep=endash, labelfont=bf, justification=raggedright, width=1.33\linewidth]{caption}

\usepackage[utf8]{inputenc} 
\usepackage[T1]{fontenc}    
\usepackage{hyperref}       
\usepackage{url}            
\usepackage{booktabs}       
\usepackage{amsfonts}       
\usepackage{nicefrac}       
\usepackage{microtype}      
\usepackage{lipsum}
\usepackage{graphicx}
\graphicspath{ {./images/} }

\usepackage{csquotes}

\numberwithin{equation}{section}

\DeclareMathSymbol{\sm}{\mathbin}{AMSa}{"39}

\theoremstyle{definition}
\newtheorem{defn}{Definition}[section]

\newtheorem{prop}{Proposition}[section]
\newtheorem{cor}{Corollary}[section]

\newcommand{\dt}{\,\mathrm{dt}}
\newcommand{\ds}{\,\mathrm{ds}}

\newcommand{\I}{\mathcal{I}}

\newcommand{\Cost}{\text{Cost}}

\newcommand{\R}{\mathbf{R}} 

\newcommand{\intzo}{\int_0^1}

\newcommand{\ema}{e^{-\alpha}}

\newcommand{\inv}{^{\sm 1}}

\newcommand{\alami}{a_{\lambda_i}}

\newcommand{\dotalami}{\dot{a}_{\lambda_i}}
\newcommand{\ddotalami}{\ddot{a}_{\lambda_i}}

\newcommand{\partialt}{\frac{\partial}{\partial t}}
\newcommand{\partiald}[2]{\frac{\partial #1}{\partial #2}}

\usepackage{authblk}

\title{Competitive equilibria in trading\thanks{This is version three of the paper. The major change relative to the prior version is a cleanup of the centralization plots, correction of typos and cleaned up plots. This paper also adds an expression for the optimal number of traders to represent when centralizing trades, see \eqref{eq:optimal-to-add}. There is also a new Section \ref{sec:normalizations} regarding time and target quantity normalizations which is mainly a reflection the meaning of those normalizations as they relate to market impact parameters.}}

\author{
  Neil A. Chriss\\
  \texttt{neil.chriss@gmail.com} \\
}

\begin{document}
\maketitle
\begin{abstract}
This is  the third paper in a series concerning the game-theoretic aspects of position-building while in competition. The first paper \cite{arXiv:2409.03586v1} set forth foundations and laid out the essential goal, which is to minimize implementation costs in light of how other traders are likely to trade. The majority of results in \cite{arXiv:2409.03586v1} center on the two traders in competition  and equilibrium results are presented. The second paper, \cite{chriss2024positionbuildingcompetitionrealworldconstraints}, introduces computational methods based on Fourier Series which allows the introduction of a broad range of constraints into the optimal strategies derived. The current paper returns to the unconstrained case and provides a \textit{complete} solution to finding equilibrium strategies in competition and handles completely arbitrary situations. As a result we present a detailed analysis of the value (or not) of \textit{trade centralization} and we show that firms who \textit{naively} centralize trades do not generally benefit and sometimes, in fact, lose. On the other hand, firms that \textit{strategically} centralize their trades generally will be able to benefit.
\end{abstract}


\section{Introduction}

Game theory is a framework for analyzing competitive interactions among players whose fortunes depend on the actions of all participants involved. One of its central concepts is \textit{Nash equilibrium}, the set of strategies that are simultaneously \textit{optimal} with respect all of the others. The focus of this paper is to give a complete analysis of Nash equilibrium when the strategic interactions are two or more traders simultaneously purchasing the stock of a particular company over a fixed period of time. This continues the work in \cite{arXiv:2409.03586v1} which introduced the concept of trading in competition, defined standard terminology and found Nash equilibria for two-trader general and multi-trader symmetric equilibria, The current paper completes the work of \cite{arXiv:2409.03586v1} by providing a complete characterization of multi-trader equilibria. As an application we demonstrate the ways that firms who employ multiple traders can centralize their trades and disguise their order flow to lower trading costs. The rest of this introduction provides a high-level overview of the main results of this paper.

\subsection{Competition, implementation cost and equilibrium}

\textit{What drives competition in trading and what are the traders competing for?} We examine the situation in which two or more traders simultaneously believe that there is a \textit{catalyst} that at some nearby future time will cause the price of a stock to rise to a new price called the \textit{target price}\footnote{This is a \textit{positive} catalyst. There are obviously negative catalysts as well and the work here works identically for those}. We will sometimes refer to these traders as the \textit{informed traders} or just "the traders" when there is no risk of confusion. The difference between the current price and the target price represents a \textit{theoretical profit} that is the motivation for traders to purchase the stock. Traders individually choose the quantity they wish to trade, called their \textit{target quantity}.Each would like to capture as much of the theoretical profit as possible. 

Though these traders represent a minority of all traders who might trade in the stock, we model them as the only traders acting on the information related to the catalyst. Therefore their collective buying represents the \textit{marginal} demand for the stock and puts upward pressure on its price. We presume all traders recognize this and therefore select the sizing and timing of their trades to minimize the price they part for the stock. Standing in the way are two forms of market impact, those effects that cause the price of the stock to move as a result of their collective trading.

The first of the two market impact effects is the persistent rise in price caused bu the collective and persistent demand for the stock from the informed traders. This \textit{persistent} buying "leaks" into the market and causes other participants to re-rate the stock's value upward. This is often referred to as \textit{permanent market impact} or \textit{alpha-decay} where in this context the former means \textit{permanent} only until the catalyst occurs and the latter refers to the degradation of theoretical profit due to the trading. For a given trader, the cost of acquiring their target position is measured as the difference between the price per share of the stock immediately before trading begins and the average price paid is called the \textit{implementation cost} of the trade.

Each trader wishes to minimize their implementation cost by choosing the timing and sizing of their trades, the totality of which is called their \textit{trading strategy}. Traders select their strategies taking into account what their adversaries' strategies are likely to be. One way to think about what "choosing a strategy" means is to imagine that each trader shifts a portion of their target quantity backward and forward in time in order to balance two opposite considerations. Shifting trades earlier in time lowers later purchase prices by getting ahead of alpha-decay, but to do so comes at the cost of higher temporary impact. Determining the optimal mix depends on all the strategies but because traders are not  coordinating wit h one another they need to select their strategies according to \textit{general} principles. A \textit{Nash equilibrium} (see \cite{nash_1950}) is the key concept in this regard.

\subsection{The market strategy and equilibrium strategies}

A Nash equilibrium is a set of strategies that simultaneously minimize every trader's implementation cost. By definition, in such an equilibrium if any trader modifies their strategy they will increase their implementation cost and therefore an equilibrium is theoretically stable in the sense that each trader has no incentive to modify their strategy. Section \ref{sec:equilibrium} proves that in any multi-trader  competition for a single stock there is a unique set of equilibrium strategies that have closed-form descriptions for both the schedule of trades as well as the implementation costs. We describe this next.

In \cite{arXiv:2409.03586v1} we did not provide any general statements about equilibrium. It turns out that in order to find these requires a new ingredient. The {\em market strategy}, discussed in Section \ref{sec:market-minimization-strategy}, is simply the aggregate of all traders strategies. Though a strategy usually means that the  sizing and timing of trades is being controlled by one individual, we nevertheless treat the aggregate trading \textit{as if} it is a strategy in the conventional sense. Treating the market strategy as an \textit{actual} strategy allows us to study its properties when its constituent strategies are in equilibrium. A key result is that in an equilibrium the market strategy is an \textit{eager} strategy, that is, it acquires stock more aggressively than a straight-line strategy that maximally spreads its trades out over time (see Section \ref{fig:shapes} for more details). 

With this in hand a number of new results become available. First, the total implementation cost of the market strategy, or equivalently, the sum of the cost to each trader, is independent of what share of the total target quantity each trader seeks is distributed. It only depends on the number of traders, the temporary impact and alpha-decay parameters. This arises because there is a highly constrained relationship between the the individual strategies and the market strategy. Not only does this allow us to provide equilibrium strategies in closed-form (see Section \ref{sec:equilibrium}) but also provide formulas for the implementation cost of each strategy, e.g., see Proposition \ref{prop:cost-of-trading-eqi}. From this we immediately see another surprising fact: each traders implementation cost only depends on their target quantity as a fraction of all the other traders.

\subsection{Centralized trading and masking a firm's number of traders}

Modern investment firms such as multi-strategy hedge funds often have multiple traders who are trading in competition for the same stock \textit{despite} working for the same firm.  We can model this situation without regard to the firm itself, that is, we can treat each trader as independent and \textit{detached} from the others. We can also ask whether the firm can consolidate its trades to its advantage, that is to effect lower implementation costs. 

In Section \ref{sec:trade-centralization} we use the equilibrium strategies and cost formulas we derive in previous sections to analyze the potential benefits of \textit{trade centralization}. The key observation with respect to centralization is that prior discussions of trading in competition assume that each trader acts completely independently with no communication or coordination with the others. However, when a firm employs a \textit{subset} of the traders in competition, it is possible for the firm to modify some aspects of the game traders are engaged in, in particular the number of traders in competition for the given stock. The firm can first centralizing all its trade and then mask the true number of traders it has an "represent" to the market that it has any number it likes by simply consolidating or splitting up its orders, never changing its order quantity.

We study this in detail in Section \ref{sec:trade-centralization} using the various results regarding equilibrium trading costs. These allow us to compute the total cost to all traders in competition and it is a curious fact (though a familiar one to game-theorists) that as the total number of traders grows, the costs of trader in equilibrium grow as well, see for example Figure \ref{fig:plot_cost_increase_over_traders_25_n1_1_n_10}. Because some number of traders work for the same firm, the firm has an opportunity to "change the game" through centralization, specifically by changing the total number of traders in competition. This itself might lead one to think that a firm should therefore simply centralize their trades and trade them as a single trader, which indeed would minimize the total number of traders "in the game". However, there is another consideration, this time that for a \textit{given trader} who is trading a relative large share of the total quantity, the fewer players there are, the greater a share of total costs that trader will incur (see Figure \ref{fig:share-cost} and the discussion there).

Therefore the must be an optimal balance with regard to centralization. The optimal number of traders to "represent" is surprisingly simple. The optimal number of traders to represent is approximately the number of traders who do not work for the firm. This is outlined in Section \ref{sec:strategic-centralization} and the result itself is \cref{eq:optimal-to-add}. Perhaps the most surprising result of all is that the optimum number of traders itself does not depend on any of the other quantities involved, such as target quantities or alpha-decay.

\section{Background information and key definitions}

In this section we review the key concepts used through this paper in detail, repeating and enriching the overview in the introduction. For the most part these concepts were introduced in \cite{arXiv:2409.03586v1}, however several, including the market strategy, are new.

\subsection{Catalysts, urgency and competition}

This paper concerns trading in competition so the first thing to settle is \textit{why} there is competition and how does this shape demand for a stock. Our idealized model begins with the concept of a \textit{catalyst} that will occur at a fixed future date that may cause a \textit{significant} revaluation of the stock, that is, its price may significantly rise immediately after the catalyst. This is a fairly common situation in equity markets: hedge funds and similar market participants conduct research to find catalysts and then build positions to capture the theoretical profit available from the potential price moves. There are several important features of this setup that drive its exploration from a game-theoretic point of view.

First, we assume  there is a relatively small number of {\em informed traders} who are aware of the catalyst and want to build positions in order to profit from it. These traders are collectively setting the price for the stock because their aggregate demand is creates a persistent \textit{imbalance} in demand for the stock, relative to the its supply. Because there is a fixed date on which the catalyst will occur, traders have to acquire their positions prior to the catalyst date in order to earn their desired profit. This creates \textit{urgency} because under the implicit rules of this game, traders are competing with other trades for liquidity and have limited flexibility to modify how they trade in light of the demand imbalances they collectively create. Thus traders need a \textit{strategy} for how they are going to build their position that takes into account the likely strategies of the other traders. Thus to discuss this further we will need pin down exactly what we mean by "trading strategy".

More formally, we say the number of informed traders (henceforth, just traders) is $n$ and identify each trader by $i=1, \dots, n$ and refer to the trader whose index is $i$ as the trader $i$.  Each trader has a target quantity of stock they wish to acquire and we write $\lambda_i$ for the portion of the total quantity their target quantity 
represents. Each trader's strategy is represented by $\lambda_i a_i$ where $a_i$ is a function that says how much cumulative quantity the trader will have acquired by time $t$. This is formally a \textit{trading strategy} and we discuss this next.

\subsection{Trading strategies}
\label{sec:trading-strategies}

We study an \textit{idealized} situation and assume there are $n$ traders seeking to buy the same stock starting simultaneously at a time designated as time $t=0$. The traders are referenced by the indexes $i=1, \dots, n$. All measure their cost of acquiring the stock relative to the prevailing price at $t=0$. This price is called the \textit{reference price} or the \textit{arrival price}. At some time prior to $t=0$ each trader decides how much stock to acquire, their \textit{target quantity}. We write $\lambda_i$ for trader $i$'s target quantity and normalize their total so that $\sum_{i=1}^n \lambda_i=1$, and refer to the $\lambda_i$'s as \textit{fractional target quantities}.

At the start of trading we regard each trader's intention to buy the stock as a single order to buy, called the \textit{parent order}. This will be broken up into a number of smaller orders, the \textit{child orders}, over time. The details of how a parent order is broken up into child orders is called the \textit{trading strategy}. We formalize these ideas as follows:

\begin{itemize}
    \item Trading strategies always begin at a time denoted $t=0$ and end at time $t=1$. Thus time is {\em scaled}, this makes some of the mathematics significantly more straightforward and sacrifices no generality; 

    \item A trading strategy is called a {\em position-building} strategy if it starts with no stock at $t=0$ and ends with a {\em positive} quantity of stock (at $t=1$). We will generally discuss position-building strategies in this paper and therefore we will use the term \textit{trading strategy} to mean \textit{position-building strategy} unless otherwise stated; 

    \item A position-building strategy whose $a_i(t)$ such that $a_i(1)=1$ is called a \textit{unit strategy}; 

    \item We will assume each trader (say $i$) has a fixed quantity of the stock they wish to acquired called the \textit{target quantity}. After we normalize the sum of all target quantities to one, we will denote trader $i$'s target quantity by $\lambda_i$ and refer to it as the \textit{fractional target quantity};
    
    \item When two or more traders are building a position in the same stock over the same interval we say they are building the positions {\em in competition}. This is the primary focus here and in \cite{arXiv:2409.03586v1};

    \item The activity of multiple traders competing for the same stock will be referred to as \textit{the trade}, and each is said to be \textit{in the trade} or \textit{taking up the trade}; and
    
    \item Each trading strategy is specified by two components: a \textit{ fractional target quantity} $\lambda_i$ and a \textit{unit strategy} $a_i$ where $i$ is the index of the trader.  
\end{itemize}

Thus a trading strategy is a strategy for acquiring a target quantity of stock starting at a fixed date and ending with a desired quantity of stock prior to the catalyst date. In fact, it is a precise schedule of trades that says how much to trade at each point in time, which, naturally is a key limitation of the framework here and \cite{arXiv:2409.03586v1}. As set forth the strategies do not anticipate incorporating market information into the model as time evolves. Therefore the strategies obtained here are most useful in precisely those situations in which the basic contours of trading are not likely to change during the course of trading, which we argue is often the case. 

\subsection{A taxonomy of trading strategies}
\label{sec:shapes}

Since every trading strategy is represented by a function of time, it is possible to categorize strategies by the \textit{shape} of their graphs. Strategy shapes were discussed in \cite{arXiv:2409.03586v1} but for completeness we review the key points here. There are three categories of \textit{simple} strategies, that is, strategies whose second derivatives have constant sign: \textit{risk-averse}, \textit{risk-neutral} and \textit{eager} according to whether their second derivative is positive, zero or negative. Figure \ref{fig:shapes} gives a visual display of these:

\begin{figure}[H]
    \centering
    \includegraphics[width=0.5\linewidth]{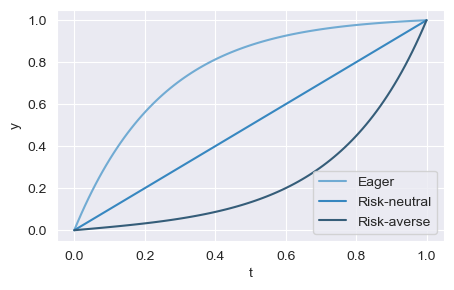}
    \caption{The three simple strategy shapes when second derivative is constant.}
    \label{fig:shapes}
\end{figure}

\begin{figure}[H]
    \centering
    \includegraphics[width=0.5\linewidth]{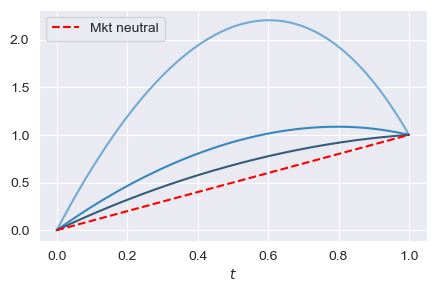}
    \caption{Aggressive strategies, only one of which is simple in that it never overbuys and therefore never sells, while the other two do.}
    \label{fig:aggressive}
\end{figure}

Eager strategies acquire more stock more rapidly than the market-neutral strategies and do so in order to \textit{front-run} other strategies, meaning, in effect, to avoid the permanent market impact of their competitors (see Section \ref{sec:implementation-cst}. Some eager strategies will \textit{overbuy} and at an interim time begin to sell. We give examples of eager strategies that overbuy, which we will call \textit{aggressive}, in Figure\ref{fig:aggressive}.

In the simplest terms, front running occurs when a trader uses advance knowledge of an upcoming large transaction to exploit market movements to their advantage. In essence, the front runner anticipates the inevitable market reaction to a large order, when the rest of the market adjusts. From a game-theoretic perspective, front running can be seen as a strategy that seeks to exploit a trader's knowledge of what other traders are likely to do in order to adjust their strategies accordingly.

\subsection{Market impact and implementation cost} 
\label{sec:implementation-cst}

Here and throughout, we assume as usual that there is a small number of informed traders competing to buy a stock. We assume that each trader has selected a strategy and ask, \textit{what will it cost each trader to acquire their position?} This cost has a specific meaning here. It is the price in excess of the reference price, that is, the price paid (per share) in excess of the prevailing price prior to the start of trading. At work are two forces, one relating to the transient impact of liquidity demand and the other related to the collective information content of the trades. The first is based on the assumption that at any given moment the aggregate demand for liquidity in the stock caused by the traders may outstrip available supply causing traders to pay a premium to have their trades executed. The second is based on the idea that when traders trade toward a catalyst the market is able to spot to some degree this activity and extract some information value from it which causes the price to persistently rise as a consequence of the trading. We work out the details of this now. 

\vskip 10pt

\textit{Trading strategies, derivatives and the market strategy}

We use the same temporary and permanent impact models as in \cite{almgren1997optimal}, \cite{almgren2001optimal} and \cite{arXiv:2409.03586v1}. In brief, temporary market impact is the immediate and \textit{incremental} cost of trading due to transient, liquidity driven effects. This is modeled as having a per unit impact on price \textit{proportional} to the \textit{net rate} of trading, where \textit{net} refers to the total over all traders trading in competition. Permanent impact, on the other hand, is an effect due to the cumulative level of trading over the life of a trade and this is modeled as having a per unit impact on price proportional to the cumulative quantity of stock acquired throughout the life of a trade. To explore these further we need to define precisely what we mean by \textit{rate of trading}.

To begin, we lay out our standard notation and setup. There are $n$ traders in competition identified by indexes $i=1, \dots, n$. We write $\lambda_i$ for the \textit{fractional target quantity} of trader $i$. This is the fraction of the total target quantity over all traders that trader $i$ will trade so that $\sum_{i=1}^n \lambda_i = 1$ and therefore total quantity is normalized to one. We write $a_i(t)$ for the \textit{unit part} of $i$'s trading strategy, and therefore $a_i(t)=0, a_i(1)=1$ and $\lambda_i \, a_i(t)$ is formally the trader's strategy. For a given strategy we write $\dot a_i(t)$ for the derivative of the strategy with respect to time. This represents the rate of trading in the stock. It is \textit{instantaneous rate of trading} and may be regarded as an \textit{instantaneous} measure of how much trader $i$ is trading at a given time. 

We also write

\begin{equation}
    m(t) = \sum_{i=1}^n \lambda_i a_i(t) 
\end{equation}

for the \textit{market strategy}. We use this anthropomorphism in the same spirit as when one says "the market rallied", despite there being no collective volition involved. We note that 

\begin{equation}
\label{eq:market-strategy-intro}
    \dot m(t) = \sum_{i=1}^n \lambda_i \dot a_i(t)
\end{equation}

is the \textit{net} rate of trading of all traders in competition. With these concepts in hand we can formally define temporary and permanent market impact.

\vskip 10pt

\textit{Temporary and permanent impact}

As usual we are concerned with $n$ traders in competition for a specific stock and we focus our attention on the trading that occurs between time $t=0$ (the start time) and $t=1$ (the completion time). Note that both time and trading quantity are \textit{normalized} to one unit, which means in effect that we have \textit{erased} all of the details concerning the level of trading relative to the level of available liquidity. We will take this point up later in the paper. 

As already noted we are interested in each trader's \textit{implementation cost} and this arises from two effects acting in tandem on the execution price of the stock and we reiterate the fact that as in \cite{arXiv:2409.03586v1} both temporary and permanent impact arise from the collective trading of all the traders in competition for the stock. In this way of looking at things, each time a trader executes a trade there are two things impacting the incremental price paid due to \textit{temporary impact:}

\begin{itemize}
    \item \textit{The trader's rate of trading:} for, say, trader $i$, the rate of trading at time $t$ is given by $\lambda_i \dot a_i(t)$. This is effectively the quantity of traded over an infinitesimal duration of time. One way to see this is to note that integral of $\lambda_i \, \dot a_i(t)$ over the interval $[t, t+\Delta]$ is precisely the change in quantity acquired over that interval. 
    $$
        \int_t^{t+\Delta t} \lambda_i \,\dot a_i(t)\dt  = \lambda_i \,\left(a(t+\Delta t) - a(t)\right) 
    $$
    Thus $\lambda_i\,\dot a_i(t)$ acts as a \textit{stand-in} for the quantity traded; and

    \item \textit{The aggregate (net) rate of trading for all traders who have taken up the trade:} we model the \textit{incremental} supply or demand for the stock as the trading pushing the price. We already defined the market strategy $m(t)=\sum_{i=1}^n \lambda_i a_i(t)$ and we therefore use $\dot m(t)$ to model the net quantity of traded by \textit{all} traders over a short period of time. We model the impact on the price of the stock at time $t$ as proportional to $\dot m(t)$.
\end{itemize}

The product of $\lambda_i \dot a_i(t)$ and $\dot m(t)$ is therefore the product of the infinitesimal quantity traded and the impact on price per share and this is how we model the total cost of trading due to temporary impact at a given time $t$.

\begin{equation}
\label{eq:temp-impact}
    \text{Temporary impact cost (trader $i$, time $t$)} \sim \dot m(t) \, \lambda_i \dot a_i(t)
\end{equation}

We model the cost of trading to trader $i$ due to persistent demand for the stock in a similar way:

\begin{equation}
\label{eq:permanent-impact}
    \text{Permanent impact cost (trader $i$, time $t$)} \sim m(t)\, \lambda_i \dot a_i(t)
\end{equation}

In \cref{eq:permanent-impact} we are expressing that the cost of trading due to permanent impact is proportional to the product of the rate of trading and the cumulative quantity acquired by \textit{all} traders. This in effect says that there is a relationship between the quantity these traders acquire and how much information \textit{leaks} into the market. Since our concern is always to minimize the total cost of trading, we focus on the sum of the temporary and permanent components of market impact. We summarize this in the function $C_i(t)$ for trader $i$ defined with $\lambda_i$ implicit as follows:

\begin{equation}
\label{eq:cost-function-trader-i}
    C_i(t) =\dot m(t)\, \lambda_i \dot a_i(t) + \kappa \, \, m(t) \lambda_i \dot a_i(t)
\end{equation}

Note that we are only interested in the value of $C_i$ up to a multiplicate constant in that all of our uses for it are in cost minimization problems. In particular this means that this is {\em not} a measure of the actual cost of trading at time $t$; however, since our aim is to find \textit{cost optimal} strategies, that is, the \textit{lowest cost}, we are only interested in cost up to a constant. In particular this allows us to \textit{always} set the constant of proportionality in temporary impact to one. 

The function $C_i$ is called the \textit{instantaneous cost function} for trader $i$ and it is the object of interest for finding equilibria in trading across multiple traders. Each trader would like to minimize their cost, but since each cost function involves $m$ this meet the criteria mentioned above of requiring each trader to take into account what all other traders are likely to do. 

\subsection{Minimizing implementation cost}

The key assumption in this paper is that traders will choose their strategies to minimize their implementation costs and have to do so based on some rational notion of what other traders will do. This will lead to the notion of Nash equilibrium stated below but first we examine implementation cost in more detail. The instantaneous cost function in \cref{eq:cost-function-trader-i}, $C_i$ takes into account the trading of all traders as the sum of temporary impact and permanent impact due to alpha-decay. The {\em implementation cost} to trader $i$ is therefore the integral from $t=0$ to $t=1$ of the cost function:

\begin{equation}
\label{eq:cost-fn}
    \Cost(\lambda_i a_i, m) = \int_0^1 C_i(t) \dt =\int_0^1 \big(\dot m(t) + \kappa \, m(t)\big) \lambda_i a_i(t) \dt 
\end{equation}

Of note, \eqref{eq:cost-fn} is the total implementation cost to trader $i$ of trading in competition with traders $i=1, \dots, n$ whose market strategy is $m = \sum \lambda_i a_i$ \textit{up to a multiplicative constant}. This constant is irrelevant, however, as our purpose is to minimize costs and the primary tool we use is the \textit{calculus of variations}. Specifically the  tool we use to find cost-minimizing strategies is the \textit{Euler-Lagrange equation}. This states that a function $a_i: [0, 1]\to \R$ that minimizes \cref{eq:cost-fn} satisfies the following linear partial differential equation, called the Euler-Lagrange equation:

\begin{equation}
\label{eq:euler-lagrange}
    \partialt \partiald{C_i}{\dot a_i} - \partiald{C_i}{a_i} = 0
\end{equation}

Note that if $C_i$ satisfies this equations then so does $\zeta C_i$ for any constant $\zeta \ne 0$ and it is in this sense that we only need to know the cost functions up to a multiplicative constant. 

We now introduce the definition of Nash equilibrium (see \cite{nash_1950}) which is the fundamental object of study in this paper.

\vskip 10pt

\begin{defn}[Nash equilibrium]
\label{def:equilibrium-formal}
Consider the situation of $n$ traders in competition for the same stock each with fractional target quantities $\lambda_1, \dots, \lambda_n$. Assume that in this context the market impact parameter is $\kappa$ then an \textit{equilibrium} is a set of trading strategies $a_1, \dots a_n$ and corresponding market strategy $m=\sum_{i=1}^n \lambda_i a_i$ with cost corresponding cost functions $C_i(t) = \lambda_i\dot m(t)\,  a_i(t) + \lambda_i  \kappa \, m(t) \, \dot a_i(t)$ such that each of 

\begin{equation}
    \Cost(\lambda_i a_i, m) = \intzo C_i(t) \dt 
\end{equation}
is simultaneously minimal. $\blacksquare$
\end{defn}

\vskip 10pt

Section \ref{sec:equilibrium} gives closed-form formulas for equilibrium strategies in general which generalizes the corresponding formulae for two-trader and $n$-trader symmetric equilibria in \cite{arXiv:2409.03586v1}.

\subsection{A note on time and quantity normalization}
\label{sec:normalizations}

Trader $i$ with trading strategy $a_i(t)$ and target quantity $\lambda_i$ will incur a total implementation cost equal to the integral of $C_i(t)$:

\begin{equation}
\label{eq:cost-normalize-0} 
    \intzo \dot m(t) \lambda_i \dot a_i(t) + \kappa m(t) \lambda_i \dot a_i(t) \dt
\end{equation}

and the Euler-Lagrange equation \cref{eq:euler-lagrange} tells us that a strategy $a_i(t)$ is \textit{cost-minimizing} for any non-zero multiple $\zeta$ of $C_i$. In particular we can divide by $\lambda_i$ and write 

\begin{equation}
\label{eq:cost-normalize-1}
    \intzo \dot m(t) \, \dot a_i(t) + \kappa \, m(t) \dot a_i(t) \dt
\end{equation}

and find the path $a_i(t)$ that minimizes \cref{eq:cost-normalize-1} and this will yield the same strategy cost-minimizing strategy $a_i(t)$. Given this we make several points to try explain what the normalizations mean in this context. 

\begin{itemize}
    \item Because \cref{eq:cost-normalize-1} has the same cost-minimizing strategy as \cref{eq:cost-normalize-0}, $\dot m(t) \, \dot a_i(t) + \kappa \, m(t) a_i(t)$ is an equally valid cost function. The target quantity of each trader is still "present" but it is buried in $m(t)=\sum_{i=1}^n \lambda_i a_i(t)$; 

    \item At times we will continue to use the cost function $C_i(t)$ as opposed to the integrand in \cref{eq:cost-normalize-1} (leaving in the $\lambda_i$ terms) because some of the machinations are more intuitive in this case;

    \item We could easily have tacked a parameter $\iota$ onto temporary impact and thus made the cost function 
    \begin{equation}
    \label{eq:cost-normalize-2}
        \iota \,\dot m(t) \dot a_i(t) + \kappa \, m(t) \dot a_i(t)    
    \end{equation}
    and the Euler-Lagrange equation will continue to yield the same cost-minimizing strategy. While \cref{eq:cost-normalize-2} offers a more precise description of both the temporary and permanent impact components by highlighting that each has its own degree of responsiveness to rate of trading and total trading respectively, it is mathematically more awkward so we implicitly multiple through by $\iota\inv$ and in practice maintain a factor of 1 in the $\dot m(t) \dot a_i(t)$ term.
\end{itemize}

\vskip 10pt

\textit{The impact of time and quantity normalizations}

As all integrals are from 0 to 1 and all target quantities sum to 1, it is worth asking what may be lost in making these normalizations. From a purely mathematical standpoint, the answer is \textit{nothing}. However, in terms of the \textit{interpretation} of the terms, something is indeed lost.

To understand this, consider the following example. Suppose there are two similar trading situations both with three traders respectively trading 50\%, 30\% and 20\% of the total target quantity. However, consider the additional data that the two cases differ only in their time frames:

\begin{itemize}
    \item \textit{Case 1:} the actual time frame is 20 days (four weeks) and the actual target quantity is 10,000,000 shares; and

    \item \textit{Case 2:} the actual time frame is 2 days and (again) the actual target quantity is 10,000,000 shares.
\end{itemize}

In the models in this paper and in \cite{arXiv:2409.03586v1}, case 1 and case 2 appear to be identical. Both would normalize the span of time over which the stock is purchase to the mathematically convenient interval $[0, 1]$. Yet clearly these two cases are very different. To begin with, compressing the trading of 10,000,000 shares of stock from 20 to 2 days requires ten times the volume of stock to be traded \textit{on average} at any given time. Given the temporary impact term sets the price impact of trading to be proportional to the instantaneous rate of trading at any given time, then on an \textit{all else equal} basis, the temporary impact of trading a given strategy would be ten times greater in case 2 (two days) versus case 1 (twenty days). The question, then, is \textit{to what extent to the parameters of market impact change when time frames are compressed or dilated?}

While this is not a question for this paper, it is worth asking whether in case 2 with ten times as much trading flowing through the market, whether the temporary impact parameter would change. Perhaps it would be be 10 times greater, which when re-normalized, the implementation cost would be

\begin{equation}
    \dot m(t) \dot a_i(t) + \frac{1}{10}\kappa \, m(t) \dot a_i(t)    
\end{equation}

However, we do not know if this is correct because we also have to ask whether the level of alpha-decay expands or contracts when trading occurs over shorter durations. There are two potential considerations here. First, over a shorter period of time there is less opportunity for the market to become aware of informativeness of the trades. On the other hand, given the volume from the traders in competition is on average ten times greater, it \textit{may} it stands to reason that there is more information being conveyed on average per unit of time. How do these two forces reconcile? While we will not pursue these questions here, it is necessary to point them out to properly understand their relation to real-world trading situations. Finally, we remark that while the above two cases involved "same quantity, different time frame", a similar set of questions clearly arise when dealing with same time frame, different quantity scenarios.

\subsection{Future directions}
\label{sec:future}

To conclude this section, we give a brief discussion of various directions one might consider. This paper's primary aim is to shed light on strategy decision making while trading in competition and we believe a game-theoretic framework is the right starting point. However, it is obvious that this framework has many shortcomings that provide a multitude of potential research directions. The good news is that the historical development of game theory itself provides a useful template for how one might evolve this research.

\textit{Game-theoretic directions}

The $n$-trader equilibria developed in this paper are described formally as an {\em $n$-player, simultaneous one-shot, perfect information, non-cooperative games}. We describe each of these briefly here:

\begin{itemize}
    \item \textit{One-shot games:} players make decisions simultaneously and only once, without the opportunity for future rounds or revisions. In these games, players must consider their strategies and payoffs based on their expectations of the other players' choices.

    \item \textit{Simultaneous games:} players make their decisions at the same time without knowing the choices of the other players. In these games, each player must decide based on their beliefs or expectations about the others' actions, as no player has information about others' moves when making their own.

    \item \textit{Perfect information games:} all players have complete knowledge of the game including the strategies every other player will play. In these games, players make decisions sequentially, fully aware of all previous moves, allowing them to optimize their strategy with complete information about the current state of the game.

    \item \textit{Non-cooperative games:} players pursue their individual interests, often competing against one another, and the outcome depends on the strategic choices of all participants, leading to an equilibrium where no player can improve their payoff by unilaterally changing their strategy. The Nash equilibrium is a key concept in analyzing non-cooperative games.
\end{itemize}

Each item in this list marks a potential direction for new research by going beyond the framework in this paper. In so doing, this takes the research in a direction that tracks the development of game theory since Nash. For example, traders repeatedly trade in competition, so to what extent would modeling this as a \textit{repeated game} impact equilibrium strategies? Also, in reality traders do not "make their moves" simultaneously, but rather different traders not only start their position-building at different times, but at times they go so far as to announce their ideas in conferences, public interview and private "idea dinners" with other investors. In addition, modeling this as a perfect information game is decidedly not a perfect information game. Traders do not know what the other traders' strategies are or what moves have occurred up to a given time. 

One game theoretic direction that seems particularly important to explore is \textit{Bayesian game theory}, see \cite{harsanyi1967games}. Bayesian game theory extends classical game theory to situations with incomplete information, where players (i.e., traders) do not have complete information about the other players strategies which could include the form of the strategies themselves, target size or restrictions a trader might place on their strategy. Put differently, Bayesian game theory provides a framework for studying games where players are uncertain about other players strategies. In this setup, each player (read: trader) forms beliefs about the other players' strategies based on a common prior distribution, updating their beliefs according to Bayes' rule to maximize their expected utility (which in this case means minimizing their expected losses).

\textit{Market impact models}

In addition to new game theoretic directions, there is also the question \textit{which market impact models should one use to inform trading in competition?} After all, this work is not exclusively of academic interest. We should be asking ourselves how different market impact models impact equilibrium strategies and which ones would yield the best results in practice. At this juncture it may seem that the only correct answer is to use models that most accurately predict the impact of trading on prices, however a counterpoint to this is that since we are not, in fact, interested in predicting market impact or implementation cost itself, it is not wholly obvious that this is the right answer. One reason for this is because the details of market impact, especially \textit{alpha-decay} will vary over time and must be estimated each time. Therefore, it is worth considering robust models that prioritize minimizing the cost of errors in parameters over the specificity of the models. Despite this, it is still worth examining the how other models might change the shape of strategies in equilibrium and importantly whether or not the nature of equilibrium itself is dependent on the market impact models used.

There is an enormous academic literature on market impact. A sample of its breadth includes market-microstructure theory (the mechanisms and processes by which securities are traded and how these factors influence price formation, liquidity, and transaction costs), empirical investigations of market impact functions, to the study of \textit{optimal execution} that studies trading strategies, ordinarily in non-competitive situations. Of particular interest are\textit{ non-linear }market impact models and decay kernels that describe how the impact of trading in the near-term effects prices and how this impact dissipates over time. Many modern treatments of market impact involve price impacts that fit within a general model described as follows. We let $K$ be a function called a \textit{decay kernel} that roughly speaking describes how the impact of trading dissipates over time, $\I$ be an \textit{impact function} that describes the \textit{functional form} of the impact of the rate of trading in a stock on its price and $a$ a trading strategy. Then the change in price is modeled as:

\begin{equation}
    \textit{Cumulative impact} = \int_0^t K(t - s) \, \I(\dot a(s)) \ds
\end{equation}

This model is sufficiently general as to capture a rich array of market impact models, including the ones used here and in \cite{arXiv:2409.03586v1} (as well as the original work on optimal execution in \cite{almgren1997optimal} and \cite{almgren2001optimal}). More broadly, the above model is general enough to describe as special cases some of the key models that have been studied empirically and theoretically including \cite{gatheral2010no}, \cite{gatheral2011optimal}, \cite{gatheral2013dynamical}, \cite{obizhaeva2013optimal} and \cite{bouchaud2009markets}. Also see \cite{farmer2006market}, \cite{almgren2005direct} and \cite{zarinelli2015beyond}\footnote{This literature is, in fact, much broader than we can do justice to here, but \cite{Gatheral2010three} and \cite{webster2023handbook} are excellent references}. 

In general, it would be interesting to investigate how changing to different market impact models influences the nature of equilibrium. Does it change the uniqueness of equilibrium? A step in this direction is possible using the computational methods developed in \cite{chriss2024positionbuildingcompetitionrealworldconstraints} which develops  methods using Fourier series that make make computing optimal strategies and equilibria more computational tractable in certain situations. This paper was developed to add \textit{real-world constraints} to strategies in competition, these methods may also be applied to studying optimal trading in competition with other impact models, e.g., the model in \cite{obizhaeva2013optimal}. 

\textit{Mathematical models}

A key assumption in this paper is that the functions describing the trading strategies are \textit{twice differentiable}. This makes it possible to study equilibrium using variational methods and fairly straightforward differential equations. It also allows for very straightforward approximation methods using Fourier series as in \cite{chriss2024positionbuildingcompetitionrealworldconstraints}. This convenience, however, comes at the price of not being able to include trading strategies that are not differentiable. On the one hand this might not be so bad because we can approximate non-differentiable functions arbitrarily closely with differentiable ones. On the other hand, when it comes to \textit{discontinuities}, it is at least worth speculating whether allowing for discontinuities is necessary. Discontinuities in trading strategies represent \textit{block trades}, including privately negotiated transactions executed outside of the open market, traded to minimize its impact on the stock's price due to its size. It is natural to ask whether allowing for block trades and potentially different impact dynamics for those trades would yield different and potentially valuable insights regarding trading in competition.

\subsection{Organization of this paper}
\label{sec:organization}

The rest of this paper is organized as follows. In Section \ref{sec:market-minimization-strategy} we examine what happens when all traders' strategies, \textit{market-wide}, are consolidated into a single strategy. We do this to examine how the associated costs of a \textit{market-wide strategy} compare with those in the equilibrium where every trader acts independently to minimize their costs. After that, in Section \ref{sec:market-stat} we study the \textit{market strategy}, the aggregation of all traders' strategies treated \textit{as if} it were a single strategy. We study the key properties of the market strategy in equilibrium and show that the market strategy is always an \textit{eager} strategy. In Section \ref{sec:equilibrium} we prove that equilibrium strategies exist and in Proposition \ref{prop:equilibrium} we provide closed form solutions for those strategies. This section also derives formulas for implementation costs in equilibrium. Finally, in Section \ref{sec:trade-centralization} we examine trade centralization and study how and to what extent it can improve implementation costs for a firm that employs multiple traders. Finally, we relegate most of the proofs of purely mathematical propositions to a set of appendices at the end of this paper. 

\section{Market-wide cost minimization}
\label{sec:market-minimization-strategy}

In this section we examine what happens when $n$ traders wish to buy the same stock but rather than trading in competition their trades are aggregated into a single trade and they are implicitly \textit{forced} to trade cooperatively. Such might be the case, for example, if the traders work for the same hedge fund and all trader's trades are sent to a \textit{central trading desk} that behaves as a single \textit{monolithic} trader. The punchline is that in such a case and in the absence of any \textit{aversion to market risk} the centralized trades should be executed as a \textit{risk-neutral} strategy.
 
We examine this here so that later we can analyze the \textit{price of anarchy} referred to in the introduction. While the result is immediate if we simply treat the market strategy as a single strategy, we show here  its precise derivation as the sum of the implementation costs of each strategy. Let $C$ be the total cost of trading (function) equal to the sum of the individual costs $C_i$. In that each $C_i$ is equal to $(\dot m + \kappa m)\lambda_i \dot a_i$ we immediately see

\begin{equation}
\label{eq:collective-loss}
    C = \sum_{i=1}^n (\dot m + \kappa m) \lambda_i\dot a_i = \dot{m}^2 + \kappa m \, \dot m 
\end{equation}

This relation depends on the fact that the market impact is {\em additive}. More precisely, the relationship replies on the fact that we are modeling the market impact of collective trading to be the \textit{super-position} of the market impact of individual trading. Namely for two general strategies $x$ and $y$ the cost of trading $x$ and $y$ simultaneously is equal to the sum costs of trading each. With that said we apply the Euler-Lagrange equation to find the strategy that minimizes implementation cost.

\begin{equation}
    \frac{\partial}{\partial t}\, \frac{\partial C}{\partial \dot m} = 2\ddot m + \kappa \dot m, \qquad 
    \frac{\partial C}{\partial m} = \kappa\, \dot m 
\end{equation}

and thus $2\ddot m + \kappa \dot m - \kappa \dot m = 0$ and this implies that $\ddot m = 0$ and therefore assuming the boundary conditions that $m(0)=0$ and $m(1)=1$ we have:

\begin{equation}
    m(t) = t
\end{equation}

The meaning of this is that when a central trading desk will take orders from each trader, assemble them into a single order and then allocate trade fills \textit{back} to each trader according to the fraction of the total quantity each trader requested. With this we can immediately compute the cost of trading to the central desk. 

\vskip 15pt

\begin{prop}[Market-wide cost minimization] If a "market-wide" central trading desk minimizes aggregate costs, it will trade the risk-neutral strategy $m(t)=t$ and the implementation cost (the aggregate cost) will be the integral of $(\dot m + \kappa m) \dot m$ and since $\dot m = 1$ we have

\begin{equation}
\label{eq:market-loss-fn}
    \textit{Cost (market-wide strategy)} = \intzo 1 + \kappa t \dt =  1 + \frac{\kappa}{2}
\end{equation}

As this represents the implementation cost to a kind-of benevolent market-wide central trading desk, it also says that each trader's cost is a proportion of the aggregate cost according to their share of the target quantity. That is, if trader $i$'s target quantity is $\lambda_i$ (where $\lambda_i$ is the fraction of the total quantity being acquired), then trader $i$'s costs is $\lambda_i(1 + \frac{\kappa}{2})$. In itself this is interesting but later in Corollary \ref{cor:price-of-anarchy} be able to give a precise comparison of this cost to the total implementation cost in the case where each trader acts independently and will see that the centralization represents a potentially significant savings. $\blacksquare$
\end{prop}

\vskip 15pt

\section{The market strategy}
\label{sec:market-stat}

In this section we return to the {\em general} situation of an $n$-trader non-cooperative game. That is, we assume each of the $n$-traders is merely trying to minimize costs relative to the loss function $C_i(t)$. In this case we are able to give a complete description of the equilibrium.

\subsection{The market strategy is eager}
\label{sec:study-mkt-strat}

In this section and the next we show that the market strategy is an {\em eager} strategy in the sense of Section \ref{sec:shapes}, and use this to calculate the aggregate cost of trading (i.e., the sum of the implementation costs for all traders) has a simple, closed-form solution that is independent of how traders' target quantities are distributed.

\vskip 10pt

\begin{prop}
\label{prop:mkt-strat-eager}
Let $\lambda_1, \dots, \lambda_n$ be the scaled target quantities of $n\ge 2$ traders in competition to buy the same stock. If $a_{\lambda_1}, \dots, a_{\lambda_n}$ are a Nash equilibrium then the market strategy $m(t) = \sum_{i=1}^n \lambda_i a_i(t)$ is the strategy:

\begin{equation}
    \label{eq:market-strategy}
    m(t) = \frac{1 - e^{-\alpha t}}{1 - e^{-\alpha}}, \qquad \alpha = \kappa\frac{n-1}{n+1}
\end{equation}
\end{prop}

\vskip 10pt

This implies an \textit{almost} immediate corollary. We see that \cref{eq:market-strategy} only involves the market impact parameter $\kappa$ and the number of traders $n$. This means that the total cost of trading the market strategy (see below, \cref{eq:aggregate-trading-cost} can only depend on those parameters. Therefore, the aggregate cost of trading over all traders, that is, the sum of the cost to each trader, does not depend on the mix of target quantities among traders.  

We can see this explicitly by examining the sum of the individual cost functions. Recalling that aggregate cost is the sum over all $i$ of $C_i(t) = \big(\dot m(t) + \kappa \, m(t) \big) \lambda_i \dot a_i(t)$ and writing 

\begin{equation}
\label{eq:aggregate-trading-cost}
\begin{split}
    \text{Aggregate cost (no centralization)} &= \sum_{i=1}^n \int_0^1 C_i(t) \dt \\
            &= \sum_{i=1}^n \int_0^1 \big(\dot m(t) + \kappa \, m(t) \big) \lambda_i \dot a_i(t) \dt \\
            &= \int_0^1 \left(\dot m(t) + \kappa \, m(t)\right) \dot m(t) \dt 
\end{split}
\end{equation}

A little manipulation of leads to the following immediate corollary whose proof is presented in Section \ref{sec:proof-total-cost-trading}.

\vskip 15pt

\begin{cor}[Aggregate cost is independent of target quantities]
\label{cor:total-cost-trading}
Let $\alami$ for $i=1, \dots n$ be an equilibrium as in Proposition \ref{prop:mkt-strat-eager} and let $\Cost(\alami)$ be the implementation cost for trader $i$ with respect this equilibrium. Then the {\em aggregate for} for all the strategies is given by the expression:

\begin{equation}
\label{eq:total-cost-trading}
    \text{Aggregate cost} = \frac{\kappa \left( \frac{1}{n+1} (e^{-2\alpha} - 1) + ( 1 - e^{-\alpha}) \right)}{(1 - e^{-\alpha})^2} = \frac{\alpha}{e^\alpha - 1} + \frac{\kappa n}{n+1}
\end{equation}

where $\alpha =\kappa\frac{n-1}{n+1}$. Therefore the aggregate cost only depends on the number of traders and the market impact parameters and is linear in the market impact parameter $\kappa$. We note that in Proposition \ref{prop:cost-of-trading-eqi} we directly derive aggregate cost as the right hand-expression. $\blacksquare$
\end{cor}

An immediate consequence of Proposition \ref{cor:total-cost-trading} is that for a given value of $\kappa$, the aggregate cost has a limiting value. The total implementation cost has a much simpler form in the limit as the number of traders grows large because $\lim_{n\to \infty} \alpha = \kappa$:

\begin{equation}
\label{eq:limiting-aggregate-cost}
\begin{split}
    \lim_{n\to\infty} \,\text{Aggregate cost} &= \frac{\kappa}{e^\kappa - 1} + \kappa \\
        &= \frac{\kappa + \kappa(e^\kappa - 1)}{e^\kappa - 1}\\
        &= \frac{\kappa e^\kappa}{e^\kappa - 1} \\
        &= \frac{\kappa}{1 - e^{-\kappa}}
\end{split}
\end{equation}

Note that the finite $n$ ratio of aggregate cost with our without centralization is given by:

\begin{equation}
    \left(\frac{\alpha}{e^\alpha - 1} + \frac{\kappa n}{n+1}\right) \left(1 + \frac{\kappa}{2}\right)^{-1}
\end{equation}

This allows us to immediately compute the price of anarchy, the limiting value as $n\to \infty$ of the ratio of the aggregate cost to the market minimizing cost (see Section \ref{sec:market-minimization-strategy}). Figure \ref{fig:anarchy} display the relationship between the price of anarchy, $\kappa$ and the number of traders.

\vskip 15pt

\begin{cor}
\label{cor:price-of-anarchy}
The {\em price of anarchy} in the limit as the number of traders grows large is given by:

\begin{equation}
    \frac{\kappa}{1 - e^{-\kappa}} \left(1 + \frac{\kappa}{2}  \right)^{-1} <  2
\end{equation}
\end{cor}

\subsection{Governing differential equations for traders in competition}
\label{sec:governing-diffeqs}

Sticking with the standard scenario of $n$ traders trading an aggregate quantity of one unit and each trader $i$ trading a portion $0<\lambda_i<1$ of the aggregate target quantity, we show that if strategies if there a set of scaled strategies $\alami$ (each of which is each equal to $\lambda_i a_i$ where $a_i$ is a unit strategy), $i=1, \dots, n$ are an equilibrium, then they and the resultant market strategy $m = \sum_{i=1}^n \lambda_i a_i$ must satisfy three related differential equations:

\vskip 15pt

\begin{prop}[Governing differential equations]
\label{prop:governing-diffeqs}

When the trading strategies $\alami$ are an equilibrium, then these strategies along with the market strategy $m$ satisfy the following {\em governing differential equations:}

\begin{subequations}
\begin{align}
    \ddot a_i - \kappa \dot a_i &= -\frac{1}{\lambda_i} (\ddot m + \kappa \dot m)\label{eq:ai-diffeq}\\
    \ddot m + \kappa \dot m &= \frac{2\kappa}{n+1} \dot m \label{eq:ddotm-kappa-mdot}\\
    \ddot m + \alpha \dot m &= 0 \label{eq:market-diffeq} 
\end{align}
\end{subequations}

where, as above, $\alpha = \kappa \frac{n-1}{n+1}$ and if it satisfies $m(0)=0, m(1)=1$. An immediate consequence of this is that:

\begin{equation}
\label{eq:ddota-minus-kappa-dota}
    \ddot a_i - \kappa \dot a_i = -\frac{1}{\lambda_i}\frac{2\kappa}{n+1} \dot m 
\end{equation}
\end{prop}

\vskip 10pt

The proof Proposition \ref{prop:governing-diffeqs} is given in Section \ref{sec:proofs}. The market strategy $m(t)=\sum_{i=1}^n \lambda_i a_i(t)$ (see \cref{eq:market-strategy} and therefore substituting this into \cref{eq:ai-diffeq} conclude:

\begin{equation}
\label{eq:equilibrium-closed-form}
    \ddot a_i - \kappa \dot a_i = -\frac{1}{\lambda_i}\cdot\frac{2\kappa}{n+1}\cdot \frac{\alpha \, e^{-\alpha t}}{1 - e^{-\alpha}}
\end{equation}

This leads to our being able to give a closed form solution for the equilibrium expressions.

\section{The equilibrium strategies}
\label{sec:equilibrium}

In this section we compute general equilibrium strategies as in Definition \ref{def:equilibrium-formal}  and explore some of their distinctive properties including deriving closed-form solutions for the cost of trading in an equilibrium.

\vskip 10pt

\begin{prop}[Equilibrium strategies]
\label{prop:equilibrium}

Suppose there are $n$ traders labeled $i=1, \dots, n$ trading in competition and that trader $i$'s fractional target quantity is $\lambda_i$. Then the strategies $a_1, \dots a_n$ defined directly below represent an equilibrium with respect to the cost functions $C_1, \dots, C_n$ defined in \cref{eq:cost-function-trader-i}:

\begin{equation}
\label{eq:a_i_solution}
    a_i(t) =  B_i (e^{\kappa t} - 1) + D_i(1 -  e^{-\alpha t})
\end{equation}

with the constants $B_i, D_i$ and $\alpha$ as follows:

\begin{equation}
\label{eq:constants}
    B_i = \frac{\lambda_i n - 1}{\lambda_i n (e^\kappa-1)}, \qquad  D_i = \frac{1}{\lambda_i n (1 - e^{-\alpha})}, \qquad \alpha = \kappa \frac{n - 1}{n + 1}
\end{equation}

and further the resultant {\em market strategy} $m(t)=\sum_{i=1}^n \lambda_i a_i(t)$ derived by summing the $a_i$ in \cref{eq:a_i_solution} again yields the {\em eager strategy} given in Proposition \ref{prop:mkt-strat-eager}. In particular, this means that these strategies simultaneously minimize $\intzo C_i(t) \dt$.

In next Section \ref{eq:symmetric} we show that for the case that $\lambda_i = 1/n$ for all $i$ each trader traders a $1/n$-scaled replica of this strategy. Also in proposition \ref{prop:cost-of-trading-eqi} we will compute the cost of trading each equilibrium strategy.
\end{prop}

\subsection{Visualizing equilibria}

In this section we provide a two of visualizations for the case of three traders to get a feel for what the shape of these strategies looks like. In each plot we show strategies for four different values of $\kappa$, the level of alpha-decay and a fixed distribution of target quantities. All plots \textit{scale} the trading strategies to a normalized ending quantity of 1 so that we can compare the shapes of each strategy.

\begin{figure}[H]
    \centering
    \includegraphics[width=1\linewidth]{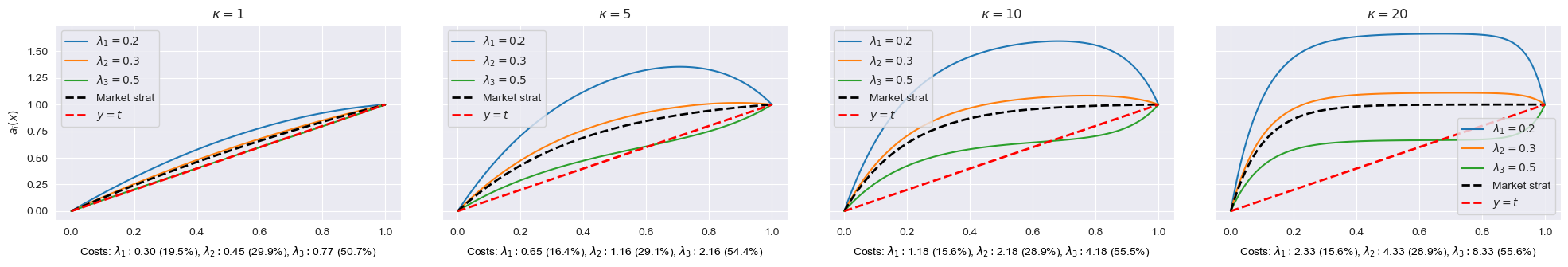}
    \caption{Each plot shows the same mix of \textit{fractional} target quantities but for different levels of $\kappa$. All plots show \textit{scaled} final quantities, for example, while all plots meet at one unit at time 1, in reality strategy one would acquire 20\% of the total. Below each plot the implementation cost to each trader is displayed along with the percentage of the total cost it represents. These are calculated using the formula for total cost given in Section \ref{sec:equi-cost-trading}. In each panel, the value of $\lambda_1$ (trader 1's percentage of total target quantity) is set to 0.2 and this corresponding strategy is in blue.}
    \label{fig:equi-strats-lambdas-var-kappa}
\end{figure}

In Figure \ref{fig:equi-strats-lambdas-var-kappa} we show three-trader equilibria for the case where traders one to three are respectively 20\%, 30\% and 50\% of the total. We see that as the level of alpha-decay increases from 1 on the left to 20 on the right, strategy 1 with 20\% of the total quantity evolves toward a "bucket" strategy, acquiring more than 100\% of its target quantity in 20\% of the time and then  selling (at a profit) toward the end to take advantage of the permanent impact due to the other two traders. We also notice that trader 2 who trades 30\% of the total quantity trades a similar strategy but to a lesser extent. Trader three trades a defensive strategy wherein it never overbuys relative to its target quantity. 

\begin{figure}[H]
    \centering
    \includegraphics[width=1\linewidth]{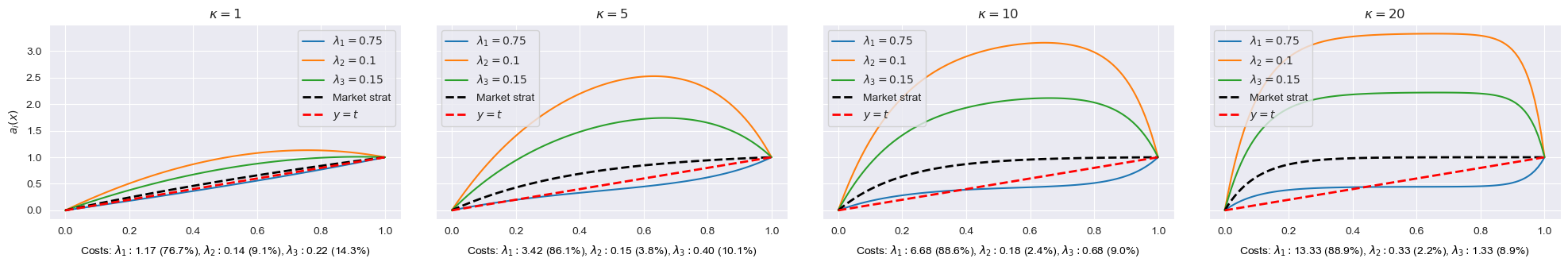}
    \caption{Each plot shows the same mix of target quantities but for different levels of $\kappa$. All plots show \textit{scaled} final quantities, for example, while all plots meet at one unit at time 1, in reality strategy one would acquire 20\% of the total. Below each plot the implementation cost to each trader is displayed along with the percentage of the total cost it represents. These are calculated using the formula for total cost given in Section \ref{sec:equi-cost-trading}. In each panel, the value of $\lambda_1$ (trader 1's percentage of total target quantity) is set to 0.2 and this corresponding strategy is in blue.}
    \label{fig:equi-strats-high-frkappa}
\end{figure}

Next in Figure \ref{fig:equi-strats-high-frkappa} we show the same mix of strategies for {\em high alpha-decay}.

\subsection{The symmetric equilibrium}
\label{eq:symmetric}

An $n$-trader symmetric equilibrium is the case where each trader trades the same quantity of stock and trades the same strategy.  This was derived in \cite{arXiv:2409.03586v1}. We now show that the general $n$-trader equilibrium contains the symmetric equilibrium as a special case, arrived at from more general principles. Thus we will show that \cref{eq:equilibrium-closed-form} is the same strategy derived in \cite{arXiv:2409.03586v1} for the special case $\lambda_i = 1/n$ and $a_i=a_j$ for all $i, j$. Plugging these values into the constants for the equilibrium strategies in \cref{eq:constants} we obtain $B_i=0$ and $D_i = \frac{1}{1 - e^{-\alpha}}$ so that the equilibrium strategy in \cref{eq:equilibrium-closed-form} becomes:

\begin{equation}
    a_i(t) = \frac{1 - e^{-\kappa\frac{n-1}{n+1}t}}{1 - e^{-\kappa \frac{n-1}{n+1}}}, \qquad \text{for all $i$}
\end{equation}

which precisely recovers the multi-trader symmetric equilibrium in \cite{arXiv:2409.03586v1}.

\subsection{The cost of trading in equilibrium}

We can use the facts we have already deduced concerning the market strategy to produce a closed-form expression for the cost of trading to each trader in an equilibrium and gather some information concerning the nature of these costs. We summarize these in the following proposition.

\vskip 15pt

\begin{prop}[The cost of trading in equilibrium]
\label{prop:cost-of-trading-eqi}
Given $n$ strategies in equilibrium with target quantities $\lambda_i$, unit strategies $a_i$ and market strategy $m = \sum \lambda_i a_i$ the cost of trading to trader $i$ $\Cost(\lambda_i a_i, m)$ is as follows:

\begin{equation}
\label{eq:equillibrium-cost-trader-i}
    \Cost(\lambda_i a_i) = \kappa \frac{\lambda_i n - 1}{n (1 - e^{-\kappa})} + \alpha \frac{1}{n(e^\alpha - 1)} + \frac{\kappa}{n+1}
\end{equation}
\end{prop}

We give a proof of this in Appendix \ref{sec:equi-cost-trading} but note a few facts here. First, the proof exploits the relationship between $\ddot a_i - \kappa \dot a_i$ and $\dot m$, without which appears to be very difficult. Second, summing $\Cost(\lambda_i a_i, m)$ over all strategies $i=1, \dots, n$ yields the total implementation cost, or simply \textit{aggregate cost}. We show in Appendix \ref{sec:equi-cost-trading} that this sum is equal to:

\begin{equation}
    \label{eq:agg-cost-2}
    \text{Aggregate cost} = \sum_{i=1}^n \Cost(\lambda_i a_i) =\frac{\alpha}{e^\alpha - 1} + \frac{\kappa n}{n+1}
\end{equation}

and while it is not immediately obvious, this is equivalent to the formula for the total implementation cost in Corollary \ref{cor:price-of-anarchy}.  

We also note that since as $n\to\infty$, $\alpha \to \kappa$, and therefore the aggregate cost expression \cref{eq:agg-cost-2} tends to $\frac{\kappa}{e^\kappa -1} + \kappa = \frac{\kappa}{1 - e^{-\kappa}}$, which is the same as \ref{eq:limiting-aggregate-cost}.  $\blacksquare$

\vskip 15pt

We summarize the most important conclusions here:

\begin{itemize}
    \item If $n$ traders in competition are in equilibrium, the there are only three variables that contribute to implementation cost: (a) the fraction of the total quantity traded by the trader (e.g., for trader $i$ this is $\lambda_i$), (b) the total number of traders, and (c) the market impact parameter.

    \item We can see from \cref{eq:equillibrium-cost-trader-i} that as a given trader's share, say $\lambda_i$, approaches 100\% of the cost, but the limit is complicated by the fact that in this case the number of traders approaches $1$. 
\end{itemize}

\subsection{Each trader's implementation cost as a share of the total cost}

We have already seen that the total implementation costs over all traders does not depend on the distribution of target quantities, a somewhat surprising fact given the fairly large range of possible strategy shapes. Nevertheless it's reasonable to ask whether each traders share of total cost is simply their {\em expected} share of the total cost. For example, if a trader is trading 25\% of the total quantity it {\em might} be the case that their share of the total cost is simply 25\% of the total. We answer this firmly in the negative in Proposition \ref{prop:imp-cost-share}.

Each trader's share of the total implementation cost is of interest because we'd like to know under what circumstances a trader is paying more or less than their {\em expected share} of total costs. The implementation cost to trader $i$ is  \cref{eq:equillibrium-cost-trader-i} and the total implementation cost is \cref{eq:agg-cost-2} and thus the ratio is: 

\begin{equation}
    S(\lambda_i a_i; \kappa, n) = \frac{\Cost(\lambda_i a_i, m)}{\text{Aggregate cost}}.
\end{equation}

We summarize the ratio's formula in the following Proposition.

\vskip 10pt

\begin{prop}
\label{prop:imp-cost-share}

Given the equilibrium strategies $\{\lambda_1 a_1, \dots, \lambda_n a_n\}$ with alpha-decay parameter $\kappa$, the share of trader $i$'s implementation cost is given by:

\begin{equation}
\label{eq:imp-cost-share}
S(\lambda_i a_i; \kappa, n) =  
         \frac{1}{n} + \frac{(n + 1) (e^\alpha - 1) }{n (1 - e^{-\kappa}) (1 - n e^\alpha)} - \frac{(n + 1) (e^\alpha - 1) }{(1 - e^{-\kappa}) (1 - n e^\alpha)} \lambda_i
\end{equation}

where as usual $\alpha = \kappa \frac{n-1}{n+1}$. This is clearly linear in $\lambda_i$ meaning that any specific trader's implementation cost as a percentage of total cost increases in proportion to their share. The proof is given in Appendix \ref{sec:imp-cost-share} and we provide some visualizations next. $\blacksquare$
\end{prop}

\subsection{Visualizations of implementation cost share}

The implementation cost share is a measure of the fraction of the total cost that a trader pays {\em relative to} what one might think of as the expected cost, namely since trader $i$'s share of the target quantity is $\lambda_i$ then one might expect that the traders share of the aggregate cost might be $\lambda_i$ as well. However this is note the case. 

\begin{figure}[H]
    \centering
    \includegraphics[width=1\linewidth]{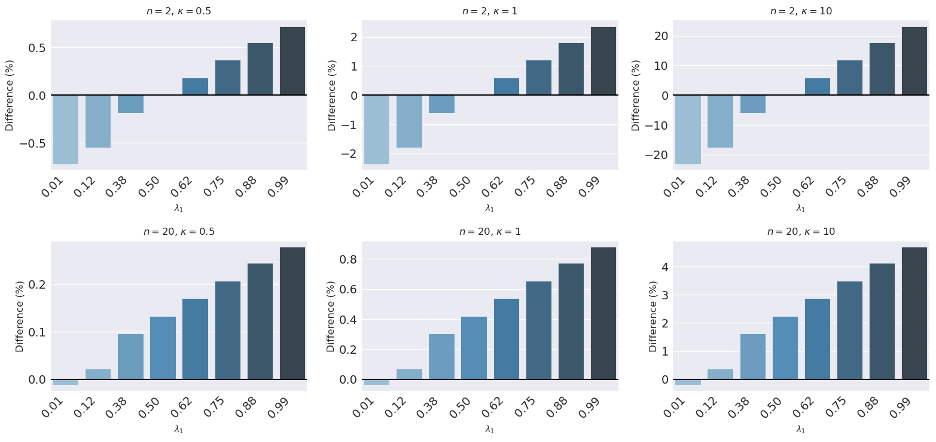}
    \caption{The deviation of a given trader's share of implementation cost from "fair share", where fair share means the same fraction of total cost as fraction of trade. For example, if trader 1 is 20\% of the trade (i.e., $\lambda_1=0.2$), then their fare share is 20\%. The bars in each plot show the change (up or down) from fair share. For example, the lower right plot shows that for traders trading 88\% and 99\% of the total target quantity, the share they actually pay is more than 4\% greater than fair share. This means that the trader trading 99\% will pay more than 100\% of total cost (and thus some traders earn a profit from the trade.}
    \label{fig:share-cost}
\end{figure}

Figure \ref{fig:share-cost} provides a visualization of this phenomena. The plots show a comparison between the \textit{actual costs} and \textit{fair-share costs}. Each plot is set up as follows:

\begin{itemize}
    \item For a trader (say 1) with a target quantity equal to $\lambda_1$, where this represents the fraction of the total quantity, the fair-share cost is $\lambda$. For example, if for a given trader $\lambda_1=0.2$ (20\% of the total quantity, then their fair-share cost is 20\%.

    \item Each plot represents a specific number of trader $n$ (e.g., the upper-left has $n=2$, two total traders) and a given $\kappa$ (e.g., the upper-left has $\kappa=0.5$). 

    \item The bars in each plot show the deviation between the fair-share cost and actual percentage of total cost in equilibrium for trader 1 for different percentage of total target quantity, denoted $\lambda_1$. For example, in the upper-left plot $\lambda_1=0.01$ means for the case where trader 1's target quantity is 1\% of the total, when there are total two traders and $\kappa=0.5$, trader 1 pays \textit{less} than its fair share by approximately 0.6\%. Meaning, roughly, that instead of 1\% of the total cost, they pay approximately 0.4\%. 
\end{itemize}

Note that in Figure \ref{fig:share-cost} the share of total cost for a given trader can exceed 100\%. For example, in the upper-right plot where $n=2$ (two traders) and $\kappa=10$ the bar for trader 1 with $\lambda_1=0.99$ is a little more than +20\%. This means that trader 1 in this case pays over 120\% of the total cost.

\section{Centralized trading}
\label{sec:trade-centralization}

We now study the question \textit{can a firm that employs a portion of all traders in competition to buy a stock lower its implementation costs by centralizing its traders?} The short answer is, \textit{it depends}, that is, the details of what precisely the firm does with the centralized trades. We characterize that details of what we mean by trade centralization:

To fix standard terminology, our setup will be that there are $n$ total traders and $n_1$ of them work for a firm, which we will referred to as simple "the firm" with traders "firm traders", while the remaining $n_2$ traders work independently and referred to as "the non-firm". As usual traders are labeled $i=1,\dots, n$ and each trader's target quantity's portion of the total is $\lambda_i$. We write $\lambda_{n_1} = \sum_{i=1}^{n_1} \lambda_i$ for the fraction of the total quantity traded by the firm, and $\lambda_{n_2}=\sum_{i=n_1+1}^{n_2} \lambda_i$ for the fraction traded by the non-firm traders:

\begin{equation}
    \underbrace{\lambda_1, \dots, \lambda_{n_1}}_{\text{Firm target qty: $\lambda_{n_1}$}} \quad \text{and} \quad \underbrace{\lambda_{n_1+1},\dots, \lambda_{n_2}}_{\text{Non-firm tgt qty: $\lambda_{n_2}$}}
\end{equation}

Each trader within the firm has a target quantity and associated parent order associated they would like to execute (see Section \ref{sec:trading-strategies}. If the firm funnels all the parent orders through an internally operated trading desk, the \textit{central trading desk}, then it is said to be \textit{centralizing its trades}. A prominent reason for considering trade centralization is that it might reduce implementation costs. The thinking behind this is that if each trader trades independently then they will operate to minimize their own implementation costs. 

As we saw in Section \ref{sec:market-minimization-strategy} the market as a whole reduces its costs relative to all traders fending for themselves. Therefore it seems likely that centralizing trades would benefit the firm through lower aggregate implementation costs. In truth, it is not so simple. It turns out that whether or not a firm can lower its costs through centralization hinges on exactly what the firm \textit{does} with the aggregated parent orders. As this is of considerable interest, we first carefully define some terms.

\begin{itemize}
    \item \textit{Firm order flow:} this is the firms detailed plan for implementing the aggregate of all firm trader's parent orders with respect to the stock they are competing to purchase;
    \item \textit{Non-cooperative trading:} this refers to the situation where all traders are trading without coordinating or communicating with others. This term is with reference to \textit{non-cooperative games} in which players are not coordinating or communicating;
    \item \textit{Cooperative trading:} when a set of traders implicitly or explicitly pool their orders and cooperate to reduce their aggregate implementation costs. This is implicitly the case when a firm centralizes its trades and seeks to minimize firm costs which is tantamount to minimizing aggregate costs;
\end{itemize}

Given this setup there are two possible approaches the firm can take. The first case is that they can allow each trader to trade independently (the \textit{no-centralization approach}) and the second is the firm can centralize all the trades (the \textit{cooperative} approach). We summarize as in the following table:

\begin{table}[H]
\centering
\begin{tabular}{|l|c|c|c|c|c|}
\hline
&  & \multicolumn{2}{c|}{\textbf{Firm}} & \multicolumn{2}{c|}{\textbf{Non-firm}} \\
 \textbf{Strategy}& \textbf{Traders} & \textbf{Traders} & \textbf{Quantity} & \textbf{Traders} & \textbf{Quantity} \\
\hline
No centralization & $n$ & $n_1$ & $\lambda_{n_1}$ & $n_2$ & $\lambda_{n_2}$ \\
\hline
Centralization & $n_2+1$ & 1& $\lambda_{n_1}$ & $n_2$ & $\lambda_{n_2}$ \\
\hline
\end{tabular}
\caption{Firm and non-firm trading setup as it pertains to how the parameters of competitive trading change when moving from non-centralized to centralized trading.}
\label{tab:coop-vs-non-coop-setup}
\end{table}

In the following sections we \textit{aggregate} the strategies in each row of Table \ref{tab:coop-vs-non-coop-setup} into two consolidated strategies: the \textit{firm strategy} and the \textit{aggregate non-firm strategy}. Like the market strategy these are not always strategies, however we treat them as such so that we can readily refer to their costs and study their properties. 

With these definitions out of the way, we distinguish between the two possible modes of centralization we will discuss. In both cases the basic setup is as in 

\begin{itemize}
    \item \textit{Naive centralization:} when a firm naively centralizes it merely takes all of the order flow and treats it as a single trade. Then the game transitions from an $n$-trader game to an $n_2+1$ trader game where $n_2$ is the number of non-firm traders and 1 represents the aggregated order flow of the firm. We call this naive because the firm trades the aggregated parent orders as one without regard to other strategic options; and
    \item \textit{Strategic centralization:} in strategic centralization the firm seeks strategic options for what \textit{to do} with the consolidated order flow. It does not necessarily trade it as a single strategy but seeks the \textit{optimal} re-organization of the flow that minimizes implementation cost.
\end{itemize}

\subsection{Naive centralization}
\label{sec:naive-centralization}

When a firm \textit{naively centralizes} it implicitly takes the view that the best way to minimize implementation costs is to trade the centralized order flow as a single monolithic strategy designed to minimize its total implementation cost. In this section, then, we study what this strategy looks like in equilibrium and compare its implementation cost to that of the de-centralized strategies (also in equilibrium). The parameters of the trading game immediately become those of the \textit{cooperative} row of Table \ref{tab:coop-vs-non-coop-setup}. Given this, there are two \textit{instances} of the firm strategy: one when the traders trade independently and the other where their trades are centralized. We refer to these as the \textit{non-centralized} and \textit{centralized} strategies respectively.

\begin{subequations}
\begin{align}
    \sum_{i=1}^{n_1} \lambda_i a_i &= \sum_{i=1}^{n_1} \lambda_i(B_i (e^{\kappa t} - 1) + D_i(1 -  e^{-\alpha t})),\qquad \text{\em Non-centralized firm strategy} \\[0.5em]
    \lambda_{n_1} a_{n_1} &= \lambda_{n_1} (B_{n_1} (e^{\kappa t} - 1) + D_{n_1} (1 - e^{-\hat{\alpha} t})), \qquad \text{\em Centralized firm strategy}
\end{align}    
\end{subequations}

with the constants $B_i$ and $D_i$ as in \cref{eq:constants}:

\begin{subequations}
\begin{align}
    B_i &= \frac{\lambda_i n - 1}{\lambda_i n (e^\kappa-1)},\quad  D_i = \frac{1}{\lambda_i n (1 - e^{-\alpha})} \\[0.5em]
    B_{n_1} &= \frac{\lambda_{n_1} (n_2 + 1) - 1}{\lambda_{n_1} (n_2 + 1) (e^\kappa-1)},\quad  D_{n_1} = \frac{1}{\lambda_{n_1} (n_2 + 1) (1 - e^{-\hat{\alpha}})} 
\end{align}
\end{subequations}

and $\alpha = \kappa \frac{n-1}{n+1}$, $\hat{\alpha} = \kappa\frac{n_2}{n_2 + 1}$ and $n_1 + n_2 = n$. Next there are the corresponding strategies for the non-firm traders:

\begin{subequations}
\begin{align}
\sum_{i=n_1+1}^{n} \lambda_i a_i &= \sum_{i=n_1+1}^{n} \lambda_i\left(B_i (e^{\kappa t} - 1) + D_i(1 -  e^{-\alpha t})\right),\qquad \text{\em Non-firm strategy (no centralization)} \\[0.5em]
\sum_{i=2}^{n_2+1} \lambda_i a_i &= \sum_{i=2}^{n_2+1} \lambda_i\left(\hat{B}_i (e^{\kappa t} - 1) + \hat{D}_i(1 -  e^{-\hat{\alpha} t})\right),\qquad \text{\em Non-firm strategy (w/ centralization)}
\end{align}    
\end{subequations}

where $B_i$ and $D_i$ are as above and $\hat{B_i}$ and $\hat{D_i}$ are as follows:

\begin{equation}
    \hat{B}_i = \frac{\lambda_i (n_2+1) - 1}{\lambda_i (n_2+1) (e^\kappa-1)},\quad  \hat{D}_i = \frac{1}{\lambda_i (n_2 + 1) (1 - e^{-\alpha})} \\[0.5em]
\end{equation}

\begin{figure}[H]
    \centering
    \includegraphics[width=1\linewidth]{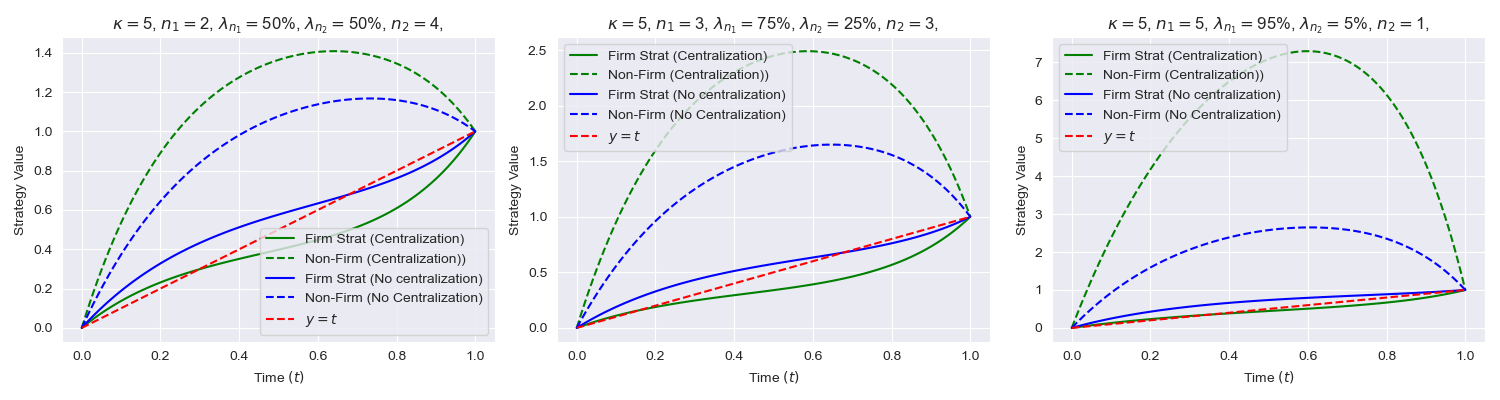}
    \caption{A comparison of aggregated firm and non-firm strategies with and without trade centralization. In this case there are six total traders, and from left to right the number of  traders in the firm is 2, 3 and 5 respectively and the firm's quantity as a fraction of the total grows from 50\% to 95\%.}
    \label{fig:strats-and-comp-strats}
\end{figure}

In Figure \ref{fig:strats-and-comp-strats} we display comparisons of the firm and non-firm strategies in the centralized versus the non-centralized cases. In each plot green lines represent equilibrium aggregated strategies when firm traders centralize their trades (the cooperative case) and blue lines represent the same but for the non-centralized case. Solid lines represent the firm's aggregated strategy in equilibrium and dashed lines represent the non-firm traders' aggregated strategy. We can think of a pair of lines of the same color (one solid and one dashed) as a strategy and a counter-strategy. We now discuss the meaning of this.

First, since dashed lines represent the non-firm traders' aggregate strategy, comparing dashed green to dashed blue lines is comparing the shape of non-firm strategies when the firm centralizes (the cooperative case, in green) versus when it doesn't. It is clear that in this case cooperation results in the non-firm traders trading more aggressively and therefore front-running to a greater degree (see Section \ref{sec:shapes}). But what happens to the firm's strategy when trades are centralized? In this case, the firm represents a greater share of the total target quantity and the total number of traders, the firm's strategy becomes generally more defensive and the result is it is more readily front-run (see Section \ref{sec:shapes}). 

\begin{figure}[H]
    \centering
    \includegraphics[width=1\linewidth]{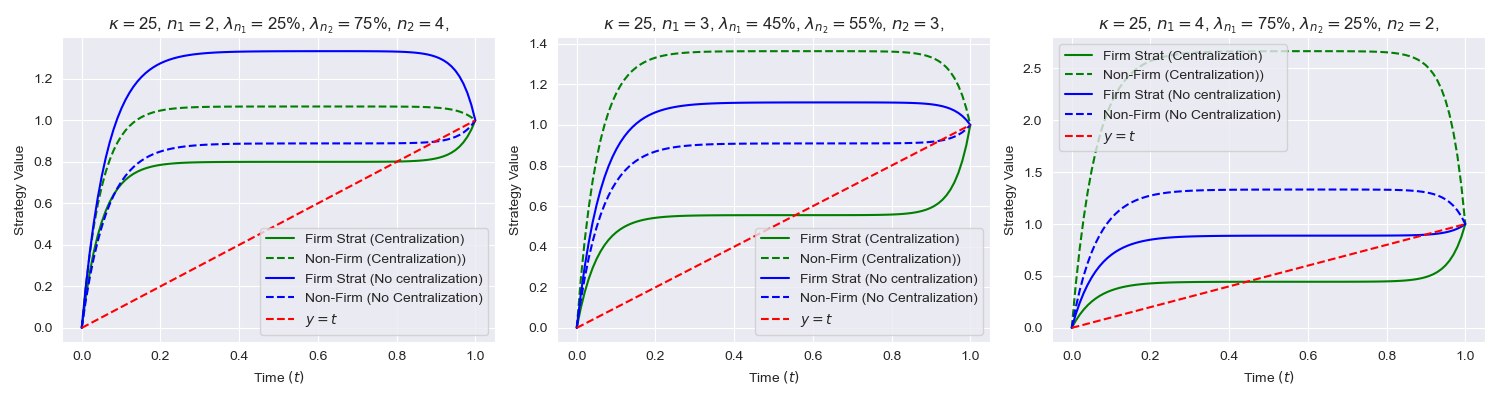}
    \caption{An example of firm versus non-firm aggregated strategies in centralized versus non-centralized cases, see Figure \ref{fig:strats-and-comp-strats}.}
    \label{fig:strats-and-comp-strats-1}
\end{figure}

Figure \ref{fig:strats-and-comp-strats-1} we provide a second example of the how centralization impacts trades.

Figures \ref{fig:strats-and-comp-strats} and \ref{fig:strats-and-comp-strats-1} suggest a counter-intuitive hypothesis. Firms that consolidate trades may not be gaining by doing so. We examine this further next.

\subsection{Naive centralization implementation costs}

Continuing with the same setup as above, we will explicitly compute the cost of each of the four strategies above, that is for the aggregated firm and non-firm strategies for the "with centralization" and "no centralization" cases in Figures \ref{fig:strats-and-comp-strats} and\ref{fig:strats-and-comp-strats-1}.

We will compare the costs of each approach. First compute the implementation costs in the case where there is \textit{no centralization}, specifically this means traders $1, \dots, n_1$ trade independently of one another.  This is nothing more than the sum of the costs in \eqref{eq:equillibrium-cost-trader-i}: 

\begin{equation}
\begin{split}
\label{eq:n1-traders-non-coop-cost}
    \textit{Firm costs (no centralization)} 
        &= \sum_{i=1}^{n_1} \Cost(\lambda_i a_i, m) \\
        &=\sum_{i=1}^{n_1}  \kappa \frac{\lambda_i n - 1}{n (1 - e^{-\kappa})} 
            + \alpha \frac{1}  {n(e^\alpha - 1)} + 
        \frac{\kappa}{n+1}\\
        &= \kappa\frac{\lambda_{n_1} n - n_1 }{n(1 - e^{-\kappa})} + \alpha \frac{n_1}{n(e^\alpha -1)} + \frac{n_1 \kappa}{n+1}
\end{split}
\end{equation}

where as usual $\alpha = \kappa\frac{n-1}{n+1}$. We can similarly compute the non-firm traders' cost in the no-centralization case:

\begin{equation}
\begin{split}
\label{eq:n2-traders-non-coop-cost}
\textit{Non-firm costs (no centralization)} &= \sum_{i=n_1+1}^{n} \Cost(\lambda_i a_i, m) \\
        &=\sum_{i=n_1 + 1}^{n}  \kappa \frac{\lambda_i n - 1}{n (1 - e^{-\kappa})} 
            + \alpha \frac{1}  {n(e^\alpha - 1)} + 
        \frac{\kappa}{n+1}\\
        &= \kappa\frac{\lambda_{n_2} n - n_2 }{n(1 - e^{-\kappa})} + \alpha \frac{n_2}{n(e^\alpha -1)} + \frac{n_2 \kappa}{n+1}
\end{split}
\end{equation}

where $\lambda_{n_2} = \sum_{i=n_1+1}^n \lambda_i$ represents the fraction of the total target quantity sought be the non-firm traders. Now we can compute the same quantities in the cooperative case. On the other hand, when the firm trades are centralized, the market-wide number of traders is $n_2+1$ consisting of the $n_2$ non-firm traders and one "consolidated" firm trader. The firm's strategy is labeled $a_{n_1}$. The cost of implementing the strategy $\lambda_{n_1}a_{n_1}$ is calculated using \cref{eq:equillibrium-cost-trader-i} setting the number of traders to $n_2+2$ and replacing $\alpha$ with $\hat{\alpha}= \kappa \frac{n_2}{n_2+2}$:

\begin{equation}
\label{eq:n1-traders-centralized-cost}
\textit{Firm cost (w/ centralization)}
    = \kappa \frac{\lambda_{n_1}(n_2 + 1)-1}{(n_2 + 1)(1 - e^{-\kappa})} + 
        \hat{\alpha} \frac{1}  {(n_2 + 1)(e^{\hat{\alpha}} - 1)} + \frac{\kappa}{n_2+2} 
\end{equation}

Next we calculate the non-firm traders' aggregate implementation cost when firm centralizes its trades. In this case the total number of traders is $n_2$ and we set $\hat{\alpha}=\kappa\frac{n_2}{n_2+2}$ as above and sum over the implementation cost of each strategy:

\begin{equation}
\begin{split}
    \textit{Non-firm costs (w/ centralization)} &= \sum_{i=n_1+1}^n \kappa \frac{\lambda_i (n_2+1) - 1}{(n_2+1)(1 - e^{-\kappa})} +  \hat{\alpha} \frac{1}{(n_2+1)(e^{\hat{\alpha}} - 1)} + \frac{\kappa}{n_2 + 2} \\
    &= \kappa \frac{\lambda_{n_2} (n_2+1) - n_2}{(n_2+1)(1 - e^{-\kappa})} +  \hat{\alpha} \frac{n_2}{(n_2+1)(e^{\hat{\alpha}} - 1)} + \frac{\kappa n_2}{n_2 + 2} 
\end{split}
\end{equation}

\subsection{Detailed numerical examples of naive centralization}
\label{sec:naive-centralization-numeric}

We now provide a detailed examination of the difference in cost when a firm centralizes its trades based on the implementation cost formulas presented above. As in the prior section there is a firm with $n_1$ traders, $n_2$ other traders and (the non-firm traders) and the total target quantity is scaled to one. The firm's portion of the target quantity is $\lambda_{n_1}$. This section provides data and analysis of the change in implementation costs to the firm and the non-firm traders in the aggregate, when the firm switches to centralized trading. Note that the general picture may seem counter-intuitive, as firm's do not always or generally experience lower implementation costs (at least as far as the models go) when they centralize, because as we will see the firm becomes a target for front-running as its trades become larger and more predictable (see Section \ref{sec:shapes}). In Section \ref{sec:strategic-centralization} we look at related and more general scenarios where a firm masks the number of traders it actually has, thus representing a different number to the market, with the express intention of reducing its implementation costs.

Tables \ref{tab:mean_values_combined} and \ref{tab:mean_values_combined-1} display the average change in implementation cost when a firm switches all of its traders from trading independently to trading through a centralized desk. The reason we say \textit{average} here is because each row in each table is an average over randomly selected values for the distribution of target quantities, number of firm traders and total number of traders. The two tables are split according to the average number of traders in the firm relative to non-firm traders. In both tables the total number of traders ($n$) is randomly selected from 20, 21 or 22, but in the first the number of firm traders ($n_1$) is randomly selected from 3, 4 or 5, while in the second it is selected from 14, 15 or 16. Each table is broken up into three blocks according to the value of $\kappa$, representing the level of alpha-decay. Each row then is an average value of the 

The big picture may be summarized as \textit{when a firm centralizes its trades the total implementation over all traders decreases but the benefits accrue to the non-firm traders and not the firm.} As this is undoubtedly a surprising result it's important to pin down precisely what the underlying assumptions are. For starters, for this to hold the market must \textit{realize} that the firm is trading as a single \textit{monolithic} block and adjust to this to form a new equilibrium. With this said, we make the following comments regarding this. When a firm \textit{naively}centralizes its trades the following happens with respect to implementation costs:

\begin{itemize}
    \item The \textit{aggregate} firm implementation cost generally \textit{increases} especially when the number of traders in the firm is a high proportion of all traders; 
    \item Overall (across all traders firm and non-firm) implementation costs decrease, consistent with the market minimization results in Section \ref{sec:market-minimization-strategy}. However, these improvements generally accrue to the non-firm traders;
    \item The implementation cost \textit{increase} to the firm is greatest \textit{on a percentage basis} when the firm is trading a small fraction of the total quantity.
\end{itemize}

\begin{table}[!ht]
\centering
\begin{tabular}{lrrrrrrrrr}
\toprule
    &  & & \multicolumn{3}{c}{\textbf{Percent Change}} & \multicolumn{2}{c}{\textbf{Costs (Firm)}} & \multicolumn{2}{c}{\textbf{Costs (Non-firm)}} \\
 $\lambda_{n_1}$ & $n_1$ & $\kappa$ & Firm & Non-firm & Total & No Central. & W/ Central. & No Central. & W/ Central. \\
\midrule
 0.07 & 4.00 & 1.00 & 1.6\% & -0.3\% & -0.1\% & 0.12 & 0.12 & 1.45 & 1.45 \\
 0.15 & 4.00 & 1.00 & 0.8\% & -0.3\% & -0.1\% & 0.23 & 0.24 & 1.33 & 1.33 \\
 0.40 & 4.00 & 1.00 & 0.3\% & -0.4\% & -0.1\% & 0.63 & 0.63 & 0.94 & 0.93 \\
 0.62 & 4.00 & 1.00 & 0.2\% & -0.7\% & -0.1\% & 0.99 & 0.99 & 0.58 & 0.58 \\
 0.82 & 4.00 & 1.00 & 0.1\% & -1.7\% & -0.1\% & 1.30 & 1.30 & 0.27 & 0.26 \\
 \midrule
 0.07 & 4.00 & 5.00 & 8.7\% & -1.3\% & -0.7\% & 0.34 & 0.36 & 4.49 & 4.43 \\
 0.15 & 4.00 & 5.00 & 4.0\% & -1.5\% & -0.7\% & 0.71 & 0.74 & 4.11 & 4.05 \\
 0.40 & 4.00 & 5.00 & 1.4\% & -2.2\% & -0.7\% & 1.97 & 2.00 & 2.85 & 2.79 \\
 0.62 & 4.00 & 5.00 & 0.9\% & -3.8\% & -0.7\% & 3.11 & 3.13 & 1.72 & 1.66 \\
 0.82 & 4.00 & 5.00 & 0.7\% & -10.1\% & -0.7\% & 4.11 & 4.14 & 0.71 & 0.65 \\
 \midrule
 0.07 & 4.00 & 25.00 & 9.5\% & -1.5\% & -0.8\% & 1.66 & 1.80 & 22.20 & 21.88 \\
 0.15 & 4.00 & 25.00 & 4.4\% & -1.6\% & -0.8\% & 3.53 & 3.68 & 20.33 & 20.00 \\
 0.40 & 4.00 & 25.00 & 1.5\% & -2.4\% & -0.8\% & 9.78 & 9.93 & 14.08 & 13.75 \\
 0.62 & 4.00 & 25.00 & 1.0\% & -4.2\% & -0.8\% & 15.41 & 15.55 & 8.45 & 8.13 \\
 0.82 & 4.00 & 25.00 & 0.7\% & -11.3\% & -0.8\% & 20.41 & 20.55 & 3.45 & 3.13 \\
 \bottomrule
\end{tabular}
\caption{The average value of change in implementation costs for a variety of scenarios when a firm switches from no centralization to centralized trading. In each case the total number of traders is one of 20, 21 and 22 and the average number of traders in the firm in this table is always 4, meaning the traders represent a minority of all traders. Each block of rows is a variety of scenarios for a fixed $\kappa$ and what varies is the fraction of the total target quantity being that the firm represents, denoted as $\lambda_1$. The percent changes columns display the change in implementation cost switching for the firm, for the aggregate of the non-firm and firm traders.}
\label{tab:mean_values_combined}
\end{table}

\begin{table}[!ht]
\centering
\begin{tabular}{lrrrrrrrrr}
\toprule
    &  & & \multicolumn{3}{c}{\textbf{Percent Change}} & \multicolumn{2}{c}{\textbf{Costs (Firm)}} & \multicolumn{2}{c}{\textbf{Costs (Non-firm)}} \\
 $\lambda_{n_1}$ & $n_1$ & $\kappa$ & Firm & Non-firm & Total & No Central. & W/ Central. & No Central. & W/ Central. \\
\midrule
 0.07 & 15.00 & 1.00 & 4.6\% & -1.8\% & -1.4\% & 0.11 & 0.11 & 1.46 & 1.43 \\
 0.15 & 15.00 & 1.00 & 2.2\% & -2.0\% & -1.4\% & 0.23 & 0.23 & 1.34 & 1.31 \\
 0.40 & 15.00 & 1.00 & 0.8\% & -2.9\% & -1.4\% & 0.62 & 0.63 & 0.95 & 0.92 \\
 0.62 & 15.00 & 1.00 & 0.5\% & -4.8\% & -1.4\% & 0.98 & 0.98 & 0.59 & 0.56 \\
 0.82 & 15.00 & 1.00 & 0.4\% & -10.9\% & -1.4\% & 1.30 & 1.30 & 0.27 & 0.25 \\
 \midrule
 0.07 & 15.00 & 5.00 & 6.8\% & -1.5\% & -1.4\% & 0.30 & 0.31 & 4.55 & 4.51 \\
 0.15 & 15.00 & 5.00 & 3.1\% & -1.7\% & -1.4\% & 0.64 & 0.66 & 4.19 & 4.15 \\
 0.40 & 15.00 & 5.00 & 1.0\% & -2.6\% & -1.4\% & 1.79 & 1.81 & 2.87 & 2.83 \\
 0.62 & 15.00 & 5.00 & 0.6\% & -4.4\% & -1.4\% & 2.95 & 2.97 & 1.73 & 1.69 \\
 0.82 & 15.00 & 5.00 & 0.4\% & -10.7\% & -1.4\% & 3.94 & 3.96 & 0.74 & 0.70 \\
 \midrule
 0.07 & 15.00 & 25.00 & 9.0\% & -1.4\% & -1.5\% & 1.57 & 1.72 & 22.22 & 21.91 \\
 0.15 & 15.00 & 25.00 & 4.2\% & -1.6\% & -1.5\% & 3.37 & 3.53 & 20.36 & 20.05 \\
 0.40 & 15.00 & 25.00 & 1.4\% & -2.3\% & -1.5\% & 9.61 & 9.77 & 14.09 & 13.78 \\
 0.62 & 15.00 & 25.00 & 0.9\% & -4.1\% & -1.5\% & 15.23 & 15.39 & 8.46 & 8.16 \\
 0.82 & 15.00 & 25.00 & 0.6\% & -11.2\% & -1.5\% & 20.22 & 20.38 & 3.46 & 3.16 \\
\bottomrule
\end{tabular}
\caption{The average value of change in implementation costs for a variety of scenarios when a firm switches from no centralized trading to centralized trading. In each case the total number of traders is one of 20, 21 and 22 and the average number of traders in the firm in this table is always 15, meaning the traders represent a majority of all traders. Each block of rows is a variety of scenarios for a fixed $\kappa$ and what varies is the fraction of the total target quantity being that the firm represents, denoted as $\lambda_1$.The percent changes columns display the change in implementation cost switching for the firm, for the aggregate of Independents and in the aggregate.}
\label{tab:mean_values_combined-1}
\end{table}

\subsection{Strategic centralization}
\label{sec:strategic-centralization}

Given the observations above, namely that centralizing trades \textit{counter-intuitively} increases firm implementation costs, it stands to reason that \textit{reversing} the process, namely of \textit{splitting trades up} in some manner might decrease firm implementation costs. Put differently, can a firm even go further and mask the true number of traders so as to \textit{deceive} the market into believing there are more traders competing than there actually are. We call this process \textit{misrepresenting order flow} and examine in what circumstances and to what extent it can improve implementation costs. We examine the change in implementation costs when a centralizes its trades and re-arranges them to make the market believe the number traders competing is different than it actually is. 

Before proceeding a small example is in order to illustrate the key points. suppose there are four traders in the firm and eight non-firm traders for a total of 12 traders taking up the trade. Then The firm has the option to \textit{naively centralize} its four traders making the apparent total number of traders taking up the trade to be nine (one from the firm and the eight non-firm traders). On the other hand, the firm, after centralizing, can split the total trades into as many trades as it wants and either route to independent brokers unaware of the total, or trade \textit{as if} they are independent. As an example, the firm could split the consolidated order into twenty independent (and much smaller) trades, each 5\% of the firm's total target quantity. There are two important points in this:

\begin{itemize}
    \item When the firm centralizes its trade to modify \textit{represent} a different number of traders than it actually has, what matters is that the total number of traders competing for the stock is modified as well. This changes the cost over all traders if trading. If the firm "adds" traders by representing more than it actually has, then total cost goes up, if it "reduces" traders, then cost goes down; and

    \item Through all of this, the firm \textit{never changes its target quantity}. The firm's target quantity is taken to be exogenously given by the aggregate of its traders.
\end{itemize}

We assume we are in the same situation as when studying trade centralization (see Section \ref{sec:trade-centralization}). There is a firm with $n_1$ traders whose total target quantity over all traders is $\lambda_{n_1}$. There are $n_2$ traders outside of the firm (the non-firm traders) whose target quantities in the aggregate are $\lambda_{n_2}$. The total number of traders is $n=n_1 + n_2$. Note that the cost to the firm with \textit{no centralization} is given by:

\begin{equation}
\label{eq:impl-cost-again}
    \textit{Firm cost (no centralization)} = \kappa\frac{\lambda_{n_1} n - n_1 }{n(1 - e^{-\kappa})} + \alpha \frac{n_1}{n(e^\alpha -1)} + \frac{n_1 \kappa}{n+1}, \quad \alpha = \kappa\frac{n-1}{n+1}
\end{equation}

Now imagine the firm centralizes its trades and divides its orders into $n_1+m$ separate \textit{parent orders}, where $n_1+ m\ge 1$. That is, the firm uses the fact that its traders are not totally autonomous to mask the true number of traders it has. This \textit{in effect} is changing the game that all traders are playing. If $m<0$ the firm is reducing the \textit{apparent} number of traders and if $m>0$ it is increasing the number of traders. Note that the firm will do this without modifying the total quantity it seeks to trade.

None of this is to say that non-firm traders know (or care) how many traders the firm has. In order to trade in equilibrium each trader needs to know the total number of traders taking up the trade and what their fractional target quantity is. Thus the firm's aim is to mask its true number of traders so as to change the total number of traders taking up the trade. In this case we have:

\begin{equation}
    n_1 \to n_1 + m, \quad n\to n + m\quad \alpha \to \hat{\alpha} = \kappa \frac{n + m - 1} {n+ m + 1}
\end{equation}

and therefore the firm's aggregate implementation cost when centralizing trades and modifying by $m$ traders is given by (after a little rearranging of  \cref{eq:impl-cost-again}):

\begin{equation}
\label{eq:add-m-traders-cost}
    \kappa \frac{\lambda_{n_1} (n+m) - (n_1 + m)}{(n+m)(1 - e^{-\kappa})} + 
    \kappa \frac{n_1+m}{n+m+1}+ 
    \hat{\alpha}\frac{n_1 + m}{(n+m) (e^{\hat{\alpha}} - 1)} 
\end{equation}

With this in hand we can provide provide insight into what a firm's optimal behavior should be concerning centralization.

\vskip 10pt

\textbf{Optimal strategic centralization:} The question is, \textit{for what $m$ is the implementation cost in \cref{eq:add-m-traders-cost} least?} If one notes that for relatively small values of  total traders $n+m$ we have $\hat{\alpha} \approx \kappa$ then we can approximate \eqref{eq:add-m-traders-cost} by

\begin{equation}
\label{eq:add-m-traders-approx}
    \kappa \frac{\lambda_{n_1} (n+m) - (n_1 + m)}{(n+m)(1 - e^{-\kappa})} + \kappa\frac{n_1 + m}{(n+m) (e^\kappa - 1)} + \kappa \frac{n_1+m}{n+m+1}
\end{equation}

and then it is straightforward to show that \cref{eq:add-m-traders-approx} the \texttt{arg-min} of \cref{eq:add-m-traders-approx} is achieved at:

\begin{equation}
\label{eq:optimal-to-add}
    m  = -n_1 + \sqrt{n_2(n_2 +1)}
\end{equation}

Put differently, the optimal number of traders for a firm to \textit{represent} to the market in order to \textit{minimize} implementation cost is given by $n_1+m$:

\begin{equation}
\label{eq:optimal-to-represent}
    m + n_1  = \sqrt{n_2(n_2 +1)}
\end{equation}

\begin{figure}[H]
    \centering
    \includegraphics[width=1\textwidth]{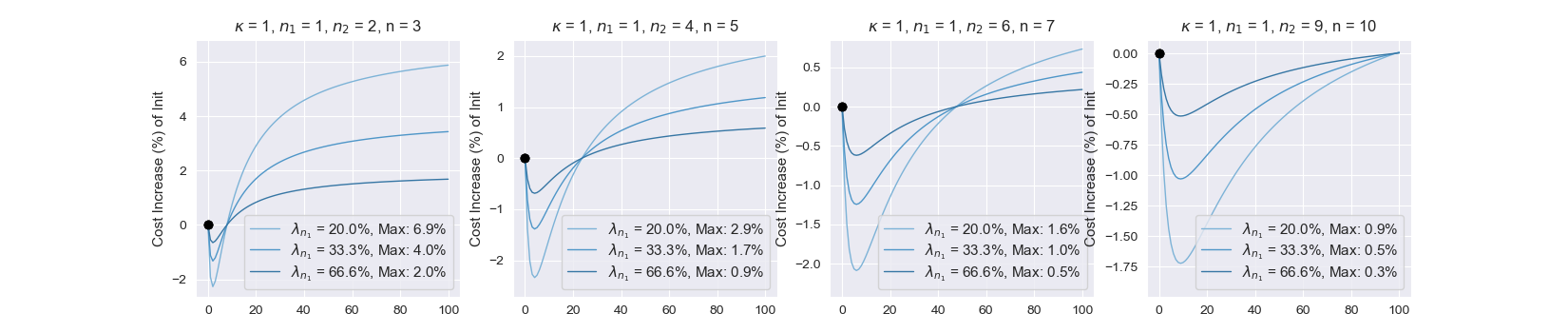}
    \caption{The firm has one trade ($n_1=1$) for relatively low alpha-decay ($\kappa=1$)for cases from left to right with non-firm traders $n_2=2, 4, 6$ and $9$. In all cases the firm's implementation cost is \textit{lower} by adding "pretending" to have more traders than it does.}
    \label{fig:plot_kappa_1_n1_1_n_10}
\end{figure}

\begin{figure}[H]
    \centering
    \includegraphics[width=1\linewidth]{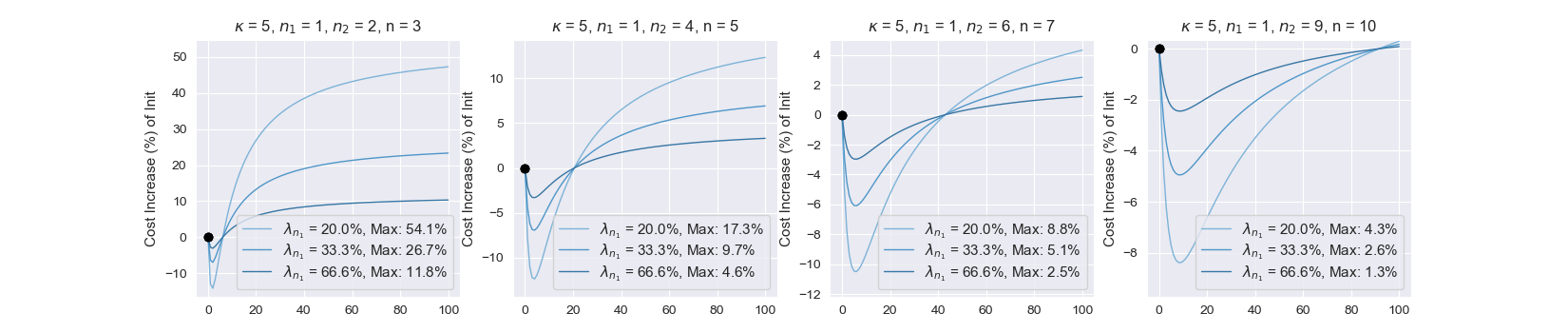}
    \caption{Identical setup as Figure \ref{fig:plot_kappa_1_n1_1_n_10} but for $\kappa=5$. The firm has one trade ($n_1=1$) for relatively low alpha-decay ($\kappa=1$)for cases from left to right with non-firm traders $n_2=2, 4, 6$ and $9$. In all cases the firm's implementation cost is \textit{lower} by adding "pretending" to have more traders than it does.}
    \label{fig:plot_kappa_5_n1_1_n_10}
\end{figure}

\begin{figure}[H]
    \centering
    \includegraphics[width=1\linewidth]{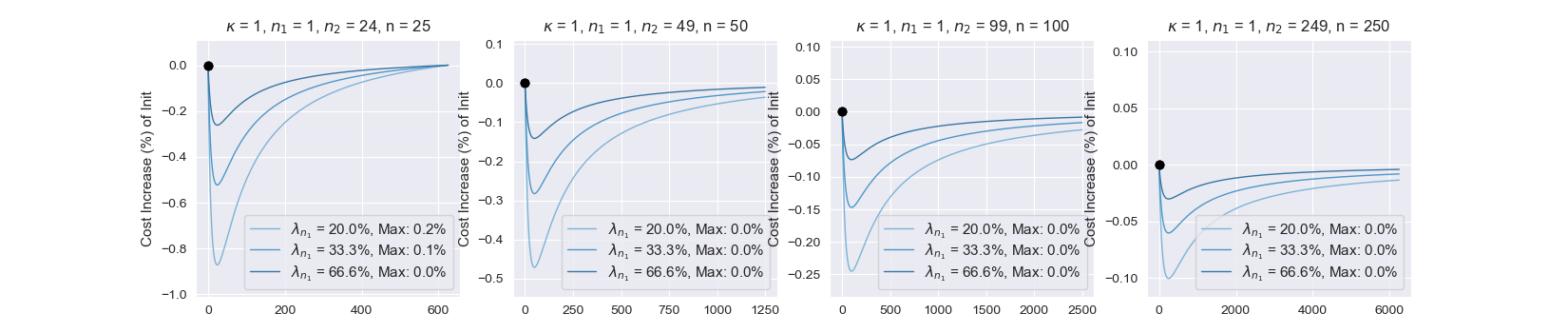}
    \caption{Similar setup as Figures \ref{fig:plot_kappa_1_n1_1_n_10} and \ref{fig:plot_kappa_5_n1_1_n_10} but this time the trade is significantly more "crowded" because the total number of traders is set to much higher values. The firm still have one trader, but now the total number of traders is from left to right is 25, 50, 100 and 250. In all cases it is still the case that adding some traders lowers costs.}
    \label{fig:plot_kappa_1_n1_1_n_250}
\end{figure}

\begin{figure}[H]
    \centering
    \includegraphics[width=1\linewidth]{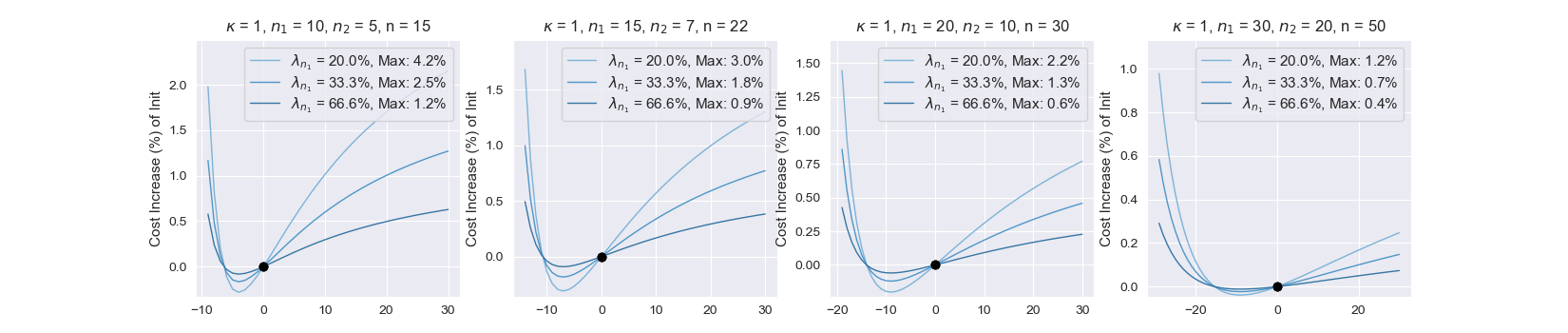}
    \caption{In this plot $\kappa=1$ so alpha-decay is relatively low. In each plot the firm has approximately 2/3rds of the total number of traders. The black dot represents the case where the firm does not centralize. One can see that consolidation (moving to the left of the black dot on the plot) lowers the cost of trading for a moderate decrease in the \textit{apparent} number of traders. We see that as the total number of traders in competition grows (as seen by the different plots starting from the left at $n=15$ versus the rightmost plot where $n=50$), the available benefit from centralization decreases.}
    \label{fig:plot_kappa_1_n1_30_n_50}
\end{figure}

We prove this in Appendix \ref{sec:sec:trade-centralization-appendix}. In other words, to optimally reduce implementation costs a firm needs to centralize its trades and then \textit{modify} its total number of traders by adding $m$ traders. Thus we conclude in most cases the optimal number of traders for the firm to represent when centralizing is between $n_2$ and $n_2 +1$. We note a few important aspects of \cref{eq:optimal-to-add}:

\begin{itemize}
    \item \textit{Balanced number of traders:} strategic centralization minimizes implementation cost when the number of traders inside the firm is set to \textit{approximately the number outside.} This is because the modification number $m \approx -n_1 + n_2$, therefore the new number of traders in the firm is $n_1 + m \approx n_2$ (one can see this visually in Figures \ref{fig:plot_kappa_1_n1_1_n_10} to \ref{fig:plot_kappa_1_n1_30_n_50};

    \item \textit{Target quantity and alpha-decay do not matter:} at least to first order, the number of trader's for the firm to represent in order to minimize implementation costs is independent of both the firm's aggregate target quantity (over all its traders) and the alpha-decay parameter $\kappa$. This is clearly the case because \cref{eq:optimal-to-add} simply has no terms involving $\lambda_{n_1}$ or $\kappa$, however we can also see this visually in Figures \ref{fig:plot_kappa_1_n1_1_n_10} to \ref{fig:plot_kappa_1_n1_30_n_50}. Finally in Appendix \ref{sec:sec:trade-centralization-appendix} one can see specifically in the derivation of the proof of \cref{eq:optimal-to-add}.
\end{itemize}

It's worth pointing out that it is somewhat surprising that the optimal number of tradings to end up with in strategic centralization is the same number of non-firm traders and that this is independent of the firm's target quantity and alpha-decay. It is also worth asking whether or not this is an artifact of features of the model itself such as the transaction cost models.

\vskip 10pt

\textbf{Limiting quantities as splitting grows:} One thing that we can do, however, is let $m\to\infty$ we see that $\hat{\alpha} \to \kappa$ and $(n_1+m)/(n+m) \to 1$  therefore the first term in \cref{eq:impl-cost-again} tends to $\frac{\kappa\lambda_{n_1} - \kappa}{1 - e^{-\kappa}}$, while the second and third terms tend to $\frac{\kappa}{e^\kappa - 1}$ and $\kappa$ respectively.  Therefore as $m\to \infty$ the firm implementation cost tends to a limiting value of:

\begin{equation}
\begin{split}
\frac{\kappa\lambda_{n_1} - \kappa}{1 - e^{-\kappa}}  + \frac{\kappa}{e^\kappa - 1} + \kappa &= \frac{\kappa\lambda_{n_1} - \kappa}{1 - e^{-\kappa}}  + \frac{\kappa e^{-\kappa}}{1 - e^{-\kappa}} +  \kappa\frac{1 - e^{-\kappa}}{1 - e^{-\kappa}} \\
    &= \frac{\kappa \lambda_{n_1}}{1 - e^{-\kappa}}
\end{split}
 \end{equation}
We can perform a similar procedure for the non-firm trader's implementation costs where $n_2$ is the number of traders \textit{not} in the firm. Here the situation is particularly simple because as the number of traders added grows, the fraction of non-firm traders goes to zero ($n_2/n \to 0$). Since the total cost for non-firm traders is given by

\begin{equation}
   \textit{Non-firm cost}  = \kappa\frac{\lambda_{n_2} n - n_2 }{n(1 - e^{-\kappa})} + \alpha \frac{n_2}{n(e^\alpha -1)} + \frac{n_2 \kappa}{n+1} 
\end{equation}

then taking the limit as $n_2/n\to 0$ yields $\frac{\kappa \lambda_{n_2}}{1 - e^{-\kappa}}$. This means that if a firm centralizes its trades and then splits them infinitely the result is that the firm and non-firm traders in the aggregate share a total cost of $\frac{\kappa}{1 - e^{-\kappa}}$ in proportion to their fractional target quantities.

\subsection{Visualizations of strategic centralization implementation costs}

We provide visual analysis of the impact of implementation costs when a firm disguises the true number of traders to make it appear \textit{as if} there are \textit{more} traders than there actually are. We adopt the conventions of the previous section and letting $n_1$ be the \textit{actual} number of firm traders, $n$ be the total number of traders. Below are a number of plots grouped into a few categories. The format is similar throughout. Each plot plots the number $m$ on the $x$-axis and the percentage change in implementation costs versus the baseline of the actual number of traders. This change is naturally the change that would occur if all traders traded a \textit{new} equilibrium based on the apparent number of traders. In other words, we assume the subterfuge works. 

All plots have three lines representing three values for the fraction the firms total quantity is relative to the total quantity, represented by $\lambda_{n_1}$=10\% (the blue lines), $\lambda_2$=33.3\%  (the orange lines) and $\lambda_3=$ 66.6\% (the green lines) of the total quantity. In cases where implementation costs \textit{rise} upon adding any number of traders, we also plot the \textit{reverse} where the number of traders is disguised to appear as if there are \textit{fewer} traders than there actually are.

\vskip 10pt

\textbf{The firm has one trader:} we make three plots (each with three sub-plots) in which the firm always has one actual trader ($n_1=1$) and the \textit{total} number of traders is 2, 3 and 10 so that the firm is 1/2, 1/3 and 1/10th of the total number of traders. We then examine implementation cost changes for three levels of alpha-decay, $\kappa=1, 2.5$ and $5$. There are a few obviously interesting trends. 

\textbf{Firm's traders represent most of the traders taking up the trade:} Below we see the case where the firm can substantially \textit{benefit} from masking its actual number of traders by \textit{consolidating} its trades so as to make it appear as if there are fewer traders in the market then there actually are.

\begin{figure}[H]
    \centering
    \includegraphics[width=1\linewidth]{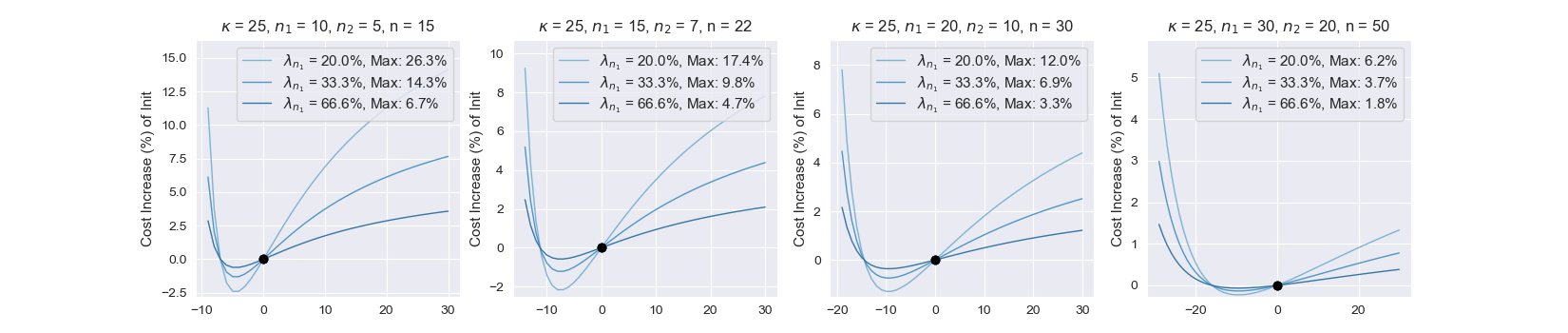}
    \caption{This is the same setup as Figure \ref{fig:plot_kappa_1_n1_30_n_50} except that alpha-decay is set to be significantly higher. In the prior plot, $\kappa$ was set to 1 and in this plot it is $25$. Generally speaking the pattern in the two plots is the same, however, in general the benefit from centralization is seen to be \textit{higher} for this case when $\kappa$ is higher.}
\end{figure}

\subsection{What is really going on with strategic centralization?}
\label{sec:what-is-going-on}

The conclusion of the prior sections is that a firm can reduce its implementation costs by centralizing and \textit{matching} the number of non-firm traders with the number of traders it represents to the market. This is the content of \cref{eq:optimal-to-add} and \cref{eq:optimal-to-represent}. As already noted it is surprising that this value is independent of both the firm's total target quantity and the alpha-decay parameter. However, our question here is, \textit{what is really going on and how is this possible at all?} First, we remind the reader that there is an optimal number of traders for a firm to represent when centralizing and this is given by \eqref{eq:optimal-to-add}. In this section we discuss \textit{why} this is the case.

The answer relates to the fact that when a firm changes the apparent number of traders in the market, they are \textit{changing the game} that all traders in competition are playing. Recall that the aggregate cost of trading in \cref{eq:agg-cost-2} is an increasing function of the number of traders $n$, while a trader's \textit{cost share} given by \ref{eq:imp-cost-share} has a complicated relationship in the number of traders $n$. We explore this now to gain intuition regarding internalization.

To start, note that as a \textit{system} the total cost of trading varies depending on the number of traders. So when a firm \textit{masks} its true number of traders it is in effect changing the game that is being played itself. Ordinarily this is not possible because each trader is totally independent of all the others. However, when there is a firm involved and that firm employs a number of traders then the firm can effectively modify the number of traders in the market to its advantage. 

Exactly what that advantage is depends on two opposing forces. The first of those forces is that as the number of traders in competition increases, all else equal the total cost of trading increases as well. This is a consequence of the degradation of efficiency a system undergoes in a non-cooperative game. The second is that for a \textit{given} trader, their share of the total cost changes when the number of traders changes. When the number of traders is very small (two or three, say) then increasing the number of traders increases share of cost for a given trader and this increase is significantly greater for traders seeking a large fraction of the total quantity. On the other hand, for a moderate number of traders, the reverse of true and increasing traders decreases cost share.

\begin{figure}[H]
    \centering
    \includegraphics[width=0.75\linewidth]{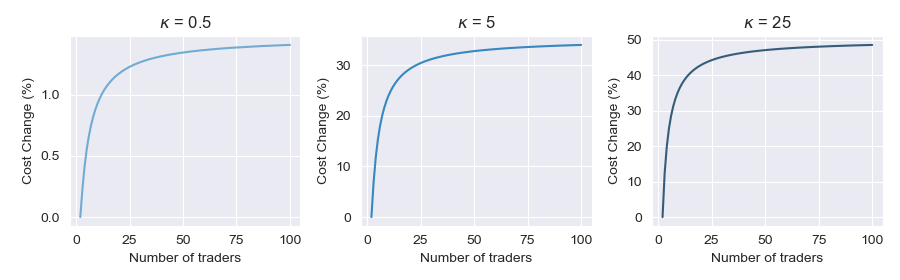}
    \caption{The relationship between the total number of traders competing for a stock and the percentage increase in the total cost of trading (over all traders) versus when there are two traders. We see that in all cases the \textit{shape} of the increase is the same and is quite sharp till approximately 20 traders. The difference between each plot, the level of alpha-decay has a significant impact on the level of the increase. We see that in the case of very high alpha decay ($\kappa=25$) costs rise to almost 50\% greater than base case.}
    \label{fig:plot_cost_increase_over_traders_25_n1_1_n_10}
\end{figure}

To see how this plays out, first observe that as the number of traders competing for a stock increases, the \textit{total cost} over all traders increases. This is another example of the general \textit{efficiency degradation} in systems where individuals act in their own interest. Figure \ref{fig:plot_cost_increase_over_traders_25_n1_1_n_10} shows the percentage change in total cost of trading versus a baseline of two traders as the number of traders overall grows.

\begin{figure}[H]
    \centering
    \includegraphics[width=0.8\linewidth]{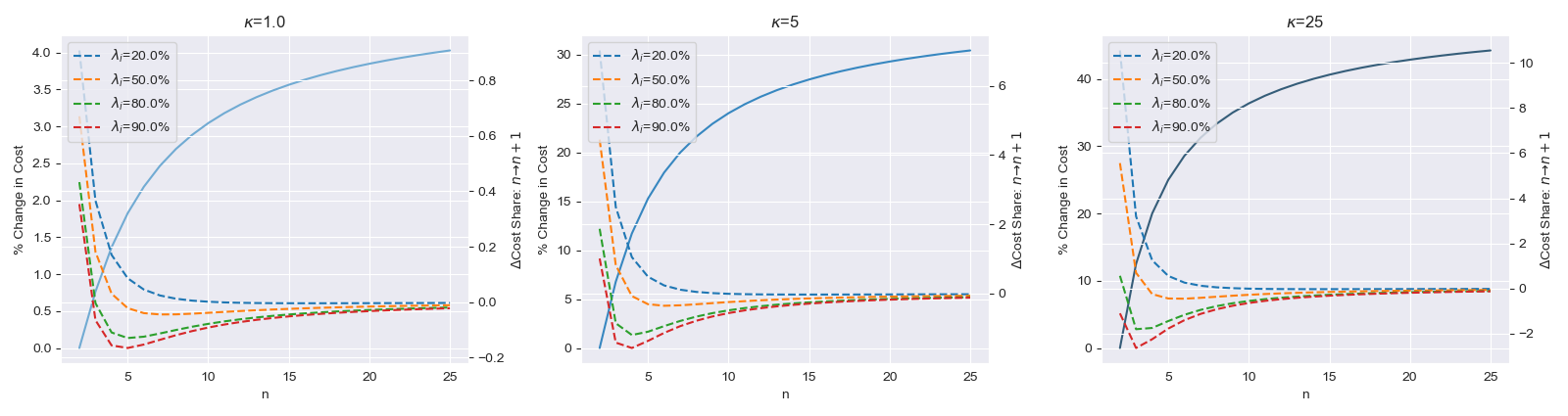}
    \caption{Each plot shows the \textit{percentage change} in total implementation cost (over all traders) as the number of traders increases from 2 to 25 where the change is relative to the two-trader case. The plots show that as more traders compete, the total cost to all traders grows and that the extent of this growth is greater as alpha-decay (given by $\kappa$) grows. For example, when $\kappa=25$, the total cost is over 40\% higher for 25 traders than 2 traders in competition. On the other hand, for a \textit{given} trader, the right-hand $y$-axis shows the change in their \textit{share} of the total cost that occurs when adding one new trader. A given point on the plot represents the change in cost share to a given trader when one more trader is added. One can see that when the number of traders is small, \textit{adding} a trader increases costs and \textit{removing} a trader therefore decreases costs.}
    \label{fig:cost-pct-change-grid}
\end{figure}

Given Figure \ref{fig:plot_cost_increase_over_traders_25_n1_1_n_10} we begin to see why there is a potential benefit to \textit{strategic centralization}. Given a starting situation, where for example a firm begins 25 traders and there are another 10 outside of the firm, then centralization allows a firm to \textit{in effect} change where the entire competition is on the graph. For example if the firm consolidates from 25 traders to a single trader (as in \textit{naive} centralization) then the firm can effectively change the total number of traders from 35 to 11, sharply decreasing the total number of traders and moving the total cost down substantially. How much impact this has overall depends on the level of $\kappa$. If this were the only force in play then naively centralizing all trades would indeed lower costs as the total cost over all traders decreases. However, the second force at play \textit{increases} the relative share of total costs to a given trader as that trader becomes a greater fraction of all traders. This is the message of Figure \ref{fig:share-cost}. Thus there is an optimal \textit{balance} to be achieved. One can see this in Figure \ref{fig:cost-pct-change-grid}.

To unpack Figure \ref{fig:cost-pct-change-grid}, note a few things about its content. 

\begin{itemize}
    \item Each plot is for a different level of alpha-decay given by $\kappa$;

    \item Each plot has plots two types of quantities as they relate to the total number of traders $n$, shown on the $x$-axis. 

    \item The solid lines are the change in the \textit{total} cost of trading relative to when there are two traders, as a function of the total number of traders using the formula \cref{eq:agg-cost-2}. For example, at $n=10$, the plot displays a little more than 3\% for the case $\kappa=1.0$, meaning that relative to 2 traders the total cost of trading for 10 traders is 3\% higher;

    \item The dashed lines are plotted on the right-hand axis and represent the change in a trader's \textit{share} of total cost that occurs in adding one more trader to the market. There are four lines in each plot representing four different traders with share of total quantity ranging from 20\% to 90\%. We can see in the right-hand plot ($\kappa=25$) that traders 3 and 4 decrease costs by 2\% when one more trader is added to the market. 

    \item Note that when the number of traders is relatively small, total cost is most rapidly increasing in number of traders, and at the same time for a moderate but not very small number of traders, increasing the number of traders \textit{decreases costs}. This is the specific reason that there is a "sweet spot" for trade centralization in terms of number of traders.
\end{itemize}

The totality of these results points to an explanation of the counter-intuitive features of trade centralization.

\section{Acknowledgments}

I extend my gratitude to the Machine Learning Research Group at Morgan Stanley, especially Yuriy Nevmyvaka for some very useful observations, and Vlad Ragulin who provided much discussion and feedback. I would also like to thank Michael Kearns and Mirah Shih of the University of Pennsylvania for extremely valuable discussions concerning algorithmic game theory.

\clearpage
\newpage

\appendix

\section{Proofs of Proposition \ref{prop:mkt-strat-eager} and \ref{prop:equilibrium} }
\label{sec:proofs}

We present a of Proposition \ref{prop:mkt-strat-eager}, which says that the market strategy is an eager strategy with a well-defined form.

\subsection{Proof of Proposition \ref{prop:mkt-strat-eager}}

To prove the equality in \cref{eq:ai-diffeq} we use the Euler-Lagrange equation on the loss function $C_i = (\dot m + \kappa m) \lambda_i a_i$. First we expand the loss function:

\begin{equation}
    C_i = \left( \sum_j \lambda_j \dot a_j + \kappa \sum \lambda_j a_j\right) \lambda_i \dot a_i
\end{equation}

and then compute the partial derivatives:

\begin{equation}
\begin{split}
    \partialt \partiald{C_i}{\dot a_i} &=  \sum_{j\ne i }^n \lambda_i \lambda_j  \ddot a_j + 2 \lambda_i^2 \ddot a_i + \kappa \sum_{j=1}^n \lambda_i\lambda_j \dot a_j   \\[0.5em]
    \partiald{C_i}{a_i} &= \kappa \lambda^2_i \dot a_i
\end{split}
\end{equation}

The Euler-Lagrange equations says that $\partialt \partiald{C_i}{\dot a_i} - \partiald{C_i}{a_i} = 0$ which implies:

\begin{equation}
    \sum_{j=1}^n \lambda_i\lambda_j \ddot a_j + \lambda_i^2 \dot a_i + \kappa \sum_{j=1}^n \lambda_i \lambda_j \dot a_j - \kappa \lambda_i^2 \dot a_i = 0
\end{equation}

Dividing everything by $\lambda_i$ substituting $\ddot m = \sum_{i=1}^n \lambda_i \ddotalami$ and $\dot m = \sum_{i=1}^n \lambda_i \dotalami$ we obtain:

\begin{equation}
    \ddot m + \lambda_i \ddot a_i + \kappa \dot m - \kappa \lambda_i \dot a_i = 0
\end{equation}

and rearranging we arrive at:

\begin{equation}
\label{eq:conclusion-proof-governing-eqs}
    \lambda_i (\ddot a_i - \kappa\dot a_i) + \ddot m + \kappa \dot m = 0 \implies
     \ddot a_i - \kappa\dot a_i = -\frac{1}{\lambda_i} (\ddot m + \kappa \dot m)
\end{equation}

and this proves the equality \cref{eq:ai-diffeq}. To complete the proof we take the weighted-sum of the right-hand-side of \cref{eq:conclusion-proof-governing-eqs} with weights equal to $\lambda_i$ to conclude:

\begin{equation}
    \ddot m - \kappa \dot m = -n (\ddot m + \kappa \dot m)
\end{equation}

rearranging this prove the equality \cref{eq:market-diffeq} $\blacksquare$

\subsection{Proof of Proposition \ref{prop:equilibrium}} 
\label{sec:proof-of-equilbrium}

We seek to show that the functions $a_i$ stated in \cref{eq:a_i_solution} satisfy \cref{eq:ai-diffeq}, which states:

\begin{equation}
    \ddot a_i - \kappa \dot a_i = -\frac{1}{\lambda_i} (\ddot m + \kappa m)
\end{equation}

To prove this we start by computing the derivatives of $a_i$:

\begin{equation}
\begin{split}
    \dot a_i &= \kappa B_i e^{\kappa t} + \frac{\alpha e^{-\alpha t}}{\lambda_i n (1 - e^{-\alpha})} \\
    \ddot a_i &= \kappa^2 B_i e^{\kappa t} - \frac{\alpha^2 e^{-\alpha t}}{\lambda_i n (1 - e^{-\alpha})} \\
\end{split}    
\end{equation}

Then forming $\ddotalami - \kappa \dotalami$ the terms with $B_i e^{\kappa t}$ cancel out and we end up with 

\begin{equation}
\begin{split}
    \ddotalami - \kappa \dotalami &= 
       -\frac{\alpha^2 e^{-\alpha t}}{\lambda_i n (1 - e^{-\alpha})} 
        - \kappa \frac{\alpha e^{-\alpha t}}{\lambda_i n (1 - e^{-\alpha})} \\
        &= -\left( \frac{\alpha^2 e^{-\alpha t}}{\lambda_i n (1 - e^{-\alpha})} 
        + \kappa \frac{\alpha e^{-\alpha t}}{\lambda_i n (1 - e^{-\alpha})}  \right) \\
        &= - \frac{\alpha e^{-\alpha t}}{\lambda_i n (1 - e^{-\alpha})} (\alpha + \kappa)
\end{split}
\end{equation}

Now note that

\begin{equation}
    \alpha + \kappa = \kappa \frac{n-1}{n+1} + \kappa = \kappa \frac{2n}{n+1}
\end{equation}

plugging this into the last line above we obtain

\begin{equation}
    \ddotalami - \kappa \lambda_i \dot a_i=  - \frac{2 \cdot \kappa \cdot \alpha e^{-\alpha t}}{\lambda_i (n+1) (1 - e^{-\alpha})} 
\end{equation}

as desired. $\blacksquare$

\subsection{Proof of the total implementation cost equations}
\label{sec:proof-total-cost-trading}

We now prove Corollary \ref{cor:total-cost-trading} which gives a closed-form formula for the aggregate cost, that is, the aggregate implementation cost. Given $n$ strategies in competition

\begin{equation}
    \int_0^1 \left(\dot m(t) + \kappa \, m(t)\right) \dot m(t) \dt
\end{equation}

where $m(t)$ is the market strategy, $\sum_{i=1}^n \alami$. Proposition \ref{prop:mkt-strat-eager} shows: 

\begin{equation}
    m(t) = \frac{1 - e^{-\alpha t}}{1 - e^{-\alpha}}
\end{equation}

and the cost of trading for each trader $i$ is $\int_0^1 \left(\dot m(t) + \kappa \, m(t)\right) \dotalami \dt $, and summing across all traders shows that the aggregate of the implementation costs over all traders:

\begin{equation}
    \text{Aggregate cost} = \int_0^1 (m(t) + \kappa \, \dot m(t)) \dot m(t) \dt 
\end{equation}

So to obtain a closed form solution we first need to compute the integrand explicitly. To start note that:

\begin{equation}
    \dot m + \kappa m = \frac{-\frac{2\kappa}{n+1} e^{-\alpha t} + \kappa }{1 - e^{-\alpha}}
\end{equation}

and $\dot m = \frac{\alpha e^{-\alpha t}}{1 - e^{-\alpha}}$ so that 

\begin{align}
    (\dot m + \kappa m) \dot m &= \frac{-\frac{2\kappa}{n+1} e^{-\alpha t} + \kappa }{(1 - e^{-\alpha})^2} \alpha e^{-\alpha t} \\[0.5em]
        &= \frac{-\frac{2\kappa \alpha}{n+1} e^{-2\alpha t} }{(1 - e^{-\alpha})^2}
            + \frac{\kappa \alpha e^{-\alpha t}}{(1 - e^{-\alpha})^2}
\end{align}

Therefore the indefinite integral is given by

\begin{equation}
    \int (\dot m + \kappa\, m) \dot m \dt = \frac{\frac{\kappa}{n+1} e^{-2\alpha t} - \kappa e^{-\alpha t}}{ (1 - e^{-\alpha})^2}
\end{equation}

and the definite integral:

\begin{equation}
\begin{split}
    \int_0^1 (\dot m + \kappa\, m) \dot m \dt &= \frac{\frac{\kappa}{n+1} e^{-2\alpha t} - \kappa e^{-\alpha t}}{ (1 - e^{-\alpha})^2} \,\Bigg |_0^1 \\
    &= \frac{\frac{\kappa}{n+1} e^{-2\alpha} - \kappa e^{-\alpha}}{ (1 - e^{-\alpha })^2} -
    \frac{\frac{\kappa}{n+1}  - \kappa }{ (1 - e^{-\alpha })^2} \\
    &= \frac{\kappa \left( \frac{1}{n+1} e^{-2\alpha} - e^{-\alpha} - \frac{1}{n+1} + 1   \right)}{(1 - e^{-\alpha})^2}
\end{split}
\end{equation}

thus

\begin{equation}
    \frac{\kappa \left( \frac{1}{n+1} (e^{-2\alpha} - 1) + ( 1 - e^{-\alpha} \right)}{(1 - e^{-\alpha})^2}
\end{equation}

However, since $e^{-2\alpha} -1 = (e^{-\alpha} - 1)(e^{-\alpha} + 1)$ the above expression simplifies to

\begin{equation}
    \frac{\frac{\kappa}{n+1} (e^{-\alpha} +1) + \kappa}{1 - \ema} 
\end{equation}

and the proposition is proved. $\blacksquare$

\subsection{Proof of equilibrium cost of trading}
\label{sec:equi-cost-trading}

In this section we compute the cost of trading to an a specific trader in an equilibrium. We first note that if $\lambda_i$, $a_i$ are equilibrium strategies with the market strategy $\sum_i \lambda_i a_i$ we have:

\begin{equation}
\label{eq:cost-of-trader-i}
    \Cost(\lambda_i a_i, m) = \lambda_i \int_0^1 (\dot m + \kappa m) \dot a_i \dt 
\end{equation}

We will use equation \cref{eq:ddota-minus-kappa-dota} that says

\begin{equation}
\label{eq:ddot}
    \kappa \dot a_i - \ddot a_i = \frac{1}{\lambda_i} \frac{2\kappa}{n+1}\dot m
\end{equation}

as follows. The cost of trading to trader $i$ is the sum of two integrals:

\begin{equation}
    \Cost(\lambda_i a_i, m) = \lambda_i \left ( I_1 + I_2 \right ), \qquad I_1 = \intzo \dot m\, \dot a_i \dt, \quad I_2 = \intzo m\, \dot a_i \dt 
\end{equation}

Using integration by parts we compute $I_1$:

\begin{equation}
    I_1 = m \, \dot a_i \Big|_0^1 - \intzo m \ddot a_i \dt \implies I_1 + I_2 = m \, \dot a_i\Big|_0^1 + \intzo m \left( \kappa \dot a_i -  \ddot a_i\right)
\end{equation}

The integrand in the last part of $I_1+I_2$ has the term $\kappa \dot a_i - \ddot a_i$ and therefore substituting \cref{eq:ddot} we obtain:

\begin{equation}
    I_1 + I_2 = m\,\dot a_i\Big|_0^1 + \frac{1}{\lambda_i} \frac{2\kappa}{n+1} \intzo m \dot m \dt 
\end{equation}

and therefore since $\Cost(\lambda_i a_i, m) = \lambda_i(I_1 + I_2)$ we have

\begin{equation}
    \Cost(\lambda_i a_i, m) = \lambda_i m \cdot \dot a_i \Big|_0^1 +  \frac{\kappa}{n+1} m^2 \Big|_0^1
\end{equation}

But because $m(t)$ is the market strategy we have $m(0)=0$ and $m(1)=1$ so we have

\begin{equation}
    \Cost(\lambda_i a_i, m) = \lambda_i \dot a_i(1) + \frac{\kappa}{n+1}
\end{equation}

Now examining the expressions for $a_i(t)$, $i=1, \dots, n$ in Proposition \ref{prop:equilibrium} we see that

\begin{equation}
    \dot a_i(1) = \kappa B_i e^\kappa + \alpha D_i e^{-\alpha}
\end{equation}

which means 

\begin{equation}
    \Cost(\lambda_i a_i, m) = \lambda_i(\kappa B_i e^\kappa + \alpha D_i e^{-\alpha}) +  \frac{\kappa}{n+1}
\end{equation}

with the constants as in \cref{eq:constants}: 

\begin{equation}
    B_i = \frac{\lambda_i n - 1}{\lambda_i n (e^\kappa-1)}, \qquad  D_i = \frac{1}{\lambda_i n (1 - e^{-\alpha})}, \qquad \alpha = \kappa \frac{n - 1}{n + 1}
\end{equation}

so that the final simplified value is 

\begin{equation}
    \Cost(\lambda_i a_i, m) = \kappa \frac{\lambda_i n - 1}{n (1 - e^{-\kappa})} +
                    \alpha \frac{1}{n(e^\alpha - 1)} + \frac{\kappa}{n+1}
\end{equation}

Note that because $\sum_{i=1}^n \lambda_i=1$, we have $\sum_{i=1}^n (\lambda_i n - 1) = n\sum_{i=1}^n \lambda_i - \sum_{i=1}^n 1 = 0$. Therefore,

\begin{equation}
    \text{Aggregate cost} = \sum \Cost(\lambda_i a_i, m) =\frac{\alpha}{e^\alpha - 1} + \frac{\kappa n}{n+1}
\end{equation}

And we note that taking the limit as the number of traders grows large yields $\alpha \to \kappa$ and that

\begin{equation}
    \lim_{n\to \infty} \text{Aggregate cost} = \frac{\kappa}{e^\kappa - 1} + \kappa
\end{equation}

And this simplifies to 

\begin{equation}
    \frac{\kappa}{e^\kappa - 1} + \frac{\kappa (e^\kappa - 1)}{e^\kappa - 1} =
        \frac{\kappa e^\kappa}{e^\kappa - 1} = \frac{\kappa}{1 - e^{-\kappa}}
\end{equation}

which recovers the same result as Corollary \ref{cor:price-of-anarchy}. $\blacksquare$

\subsection{Implementation cost share}
\label{sec:imp-cost-share}

Cost share formula can be expressed in terms of a constant term $c_0$, a linear term $c_1$ that depends on $\lambda_i$, and a denominator $d$:

\begin{equation}
\begin{split}
    c_0 &= -n \left( e^\kappa - 1 \right) \left( e^{\alpha} - 1 \right) - (n - 1) \left( e^\kappa - 1 \right) \\[0.5em]
    c_1 &= - (n + 1) n \left( e^{\alpha} - 1 \right) e^\kappa \\[0.5em]
    d &= n \left( e^\kappa - 1 \right) \left( -n \left( e^{\alpha} - 1 \right) - n + 1 \right)
\end{split}
\end{equation}

Factoring out $\left( e^\kappa - 1 \right)$, we get:

\begin{equation}
\begin{split}
    c_0 &= \left( e^\kappa - 1 \right) \left[ -n \left( e^\alpha - 1 \right) - (n - 1) \right] = \left( e^\kappa - 1 \right) \left( 1 - n e^\alpha\right)  \\[0.5em]
    c_1 &= -(n + 1) \cdot (\lambda_i \cdot n - 1) \cdot (e^\alpha - 1) \cdot e^\kappa\\
    d &= n \left( e^\kappa - 1 \right) \left[ -n \left( e^\alpha - 1 \right) - n + 1 \right] = n \left( e^\kappa - 1 \right) \left( -n e^\alpha + 1 \right).    
\end{split}
\end{equation}

Dividing $c_0$ by $d$, we obtain:

\begin{equation}
    \frac{c_0}{d} = \frac{\left( e^\kappa - 1 \right) \left( -n e^\alpha + 1 \right)}{n \left( e^\kappa - 1 \right) \left( -n e^\alpha + 1 \right)} = \frac{1}{n}.
\end{equation}

Dividing $c_1$ by $d$, we get:

\begin{equation}
    \frac{c_1}{d} = \frac{ -(n + 1) \cdot (\lambda_i \cdot n - 1) \cdot (e^\alpha - 1) \cdot e^\kappa}{n \left( e^\kappa - 1 \right) \left( -n e^\alpha + 1 \right)}
\end{equation}

and the cost share formula becomes

\begin{equation}
    \frac{c_0}{d} + \frac{c_1}{d} = \frac{1}{n} + \frac{c_1}{d}
\end{equation}

Given the expression:

\begin{equation}
\frac{c_1}{d} = \frac{-(n + 1) (\lambda_i n - 1) (e^\alpha - 1) e^\kappa}{n (e^\kappa - 1) (1 - n e^\alpha)}
\end{equation}

We want to express this as $A + B \lambda_i $. After separating and simplifying, we get:

\begin{equation}
\frac{c_1}{d} = A + B \lambda_i
\end{equation}

where:

\begin{equation}
A = \frac{(n + 1) (e^\alpha - 1) e^\kappa}{n (e^\kappa - 1) (1 - n e^\alpha)}
\end{equation}

and

\begin{equation}
B = \frac{-(n + 1) (e^\alpha - 1) e^\kappa}{(e^\kappa - 1) (1 - n e^\alpha)}
\end{equation}

and therefore the cost share formula becomes 

\begin{equation}
    \frac{1}{n} + \frac{(n + 1) (e^\alpha - 1) }{n (1 - e^{-\kappa}) (1 - n e^\alpha)} - \frac{(n + 1) (e^\alpha - 1) }{(1 - e^{-\kappa}) (1 - n e^\alpha)} \lambda_i
\end{equation}

and this proves the cost share formula. $\blacksquare$

\section{Proof of minimizing implementation costs via centralization}
\label{sec:sec:trade-centralization-appendix}

We wish to minimize the expression in \cref{eq:add-m-traders-approx} with respect to $m$. We treat $m$ as a continuous variable. To find the minimum of the expression with respect to $m$:

$$
E(m) = \kappa \frac{\lambda_{n_1} (n + m) - (n_1 + m)}{(n + m)(1 - e^{-\kappa})} + \kappa \frac{n_1 + m}{(n + m)(e^\kappa - 1)} + \frac{(n_1 + m)\kappa}{n + m + 1}
$$

we can re-parameterize it by setting $s = n + m$, so $m = s - n$. This transformation allows us to rewrite the expression in terms of $s$:

$$
E(s) = \kappa \left( \frac{\lambda_{n_1} s - (n_1 + s - n)}{s (1 - e^{-\kappa})} + \frac{n_1 + s - n}{s (e^\kappa - 1)} + \frac{n_1 + s - n}{s + 1} \right).
$$

To simplify further, let $n_2 = n - n_1$, then $n_1 + s - n = s - n_2$, and the expression becomes:

\begin{equation}
\label{eq:appendix-central-1}
E(s) = \kappa \left( \frac{(\lambda_{n_1} - 1) s + n_2}{s (1 - e^{-\kappa})} + \frac{s - n_2}{s (e^\kappa - 1)} + \frac{s - n_2}{s + 1} \right)
\end{equation}

We now take the derivative of $E(s)$ with respect to $s$ by differentiating each term in the sum individually. Writing $E_1, E_2$ and $E_3$ for the three summands in \cref{eq:appendix-central-1} we have:

\begin{subequations}
\begin{align}
    C_1(s) &= \kappa \frac{(\lambda_{n_1} - 1)s + n_2}{s (1 - e^{-\kappa})} \quad \implies \quad C_1'(s) = -\kappa \frac{n_2}{s^2 (1 - e^{-\kappa})} \\
    C_2(s) &= \kappa \frac{s - n_2}{s (e^\kappa - 1)} \quad \implies \quad C_2'(s) = \kappa \frac{n_2}{s^2 (e^\kappa - 1)} \\
   C_3(s) &= \kappa \frac{s - n_2}{s + 1} \quad \implies \quad C_3'(s) = \kappa \frac{1 + n_2}{(s + 1)^2}
\end{align}
\end{subequations}

We can see clearly that non of the three summands varies with respect to $s$ and this explains why the cost minima do not depend on number of traders added (where $m$ is number of traders added and $s=m+n$). Combining these results, we get:

\begin{equation}
C'(s) = C_1'(s) + C_2'(s) + C_3'(s) = \kappa \left( -\frac{n_2}{s^2 (1 - e^{-\kappa})} + \frac{n_2}{s^2 (e^\kappa - 1)} + \frac{1 + n_2}{(s + 1)^2} \right)    \end{equation}

We simplify the first two terms in $C'(s)$ using the identity $e^\kappa - 1 = (1 - e^{-\kappa}) e^\kappa$:

\begin{align*}
    -\frac{n_2}{s^2 (1 - e^{-\kappa})} + \frac{n_2}{s^2 (e^\kappa - 1)} &= -\frac{n_2 e^\kappa}{s^2 (e^\kappa - 1)} + \frac{n_2}{s^2 (e^\kappa - 1)} \\
        &= -\frac{n_2 e^\kappa - n_2}{s^2(e^\kappa - 1)} \\
        &= -\frac{n_2}{s^2}
\end{align*}

Thus, the derivative simplifies to:

\begin{equation}
    C'(s) = \kappa \left( -\frac{n_2}{s^2} + \frac{1 + n_2}{(s + 1)^2} \right).    
\end{equation}

We now set $C'(s) = 0$ and obtain:

\begin{equation}
    -\frac{n_2}{s^2} + \frac{1 + n_2}{(s + 1)^2} = 0.    
\end{equation}

Multiplying both sides by $s^2 (s + 1)^2$ to clear the denominators, we obtain:

\begin{equation}
    -n_2 (s + 1)^2 + (1 + n_2) s^2 = 0.    
\end{equation}

which in turn implies:

\begin{equation}
    -n_2 (s^2 + 2s + 1) + (1 + n_2) s^2 = 0,    
\end{equation}

and this simplifies to:

\begin{equation}
    s^2 - 2 n_2 s - n_2 = 0
\end{equation}

We may solve it using the quadratic formula:

\begin{equation}
    s = \frac{2 n_2 \pm \sqrt{(2 n_2)^2 + 4 n_2}}{2} = n_2 + \sqrt{n_2^2 + n_2}    
\end{equation}

Since $s = n + m$ (the original number of traders modified by $m$), it must be positive, and therefore we take the positive root:

\begin{equation}
s = n_2 + \sqrt{n_2^2 + n_2}  
\end{equation}

Finally since $s = n + m$ we have Recalling that $m = s - n$ and $n_2 = n - n_1$, we have:

\begin{equation}
    m = s - n = -n_1 + \sqrt{(n - n_1)(n - n_1 + 1)}.    
\end{equation}

and we conclude that the minimum of $E(m)$ occurs at:

\begin{equation}
m = -n_1 + \sqrt{n_2(n_2 + 1)}    
\end{equation}

and this proves the result. $\blacksquare$

\bibliographystyle{alpha}
\bibliography{competitive-equilibria-trading}

\end{document}